\documentclass[twocolumn,nofootinbib,prd,aps,amsmath,amssymb,superscriptaddress]{revtex4-1}
\usepackage{mathtools}
\usepackage{amsthm}
\usepackage{bm}% bold math
\usepackage{graphicx,epsfig}% Include figure files
\usepackage{ulem}
\usepackage{xfrac}
\usepackage{color}
\usepackage{verbatim}
\usepackage{cancel}
\usepackage{bigints}
\usepackage{caption}
\usepackage{subcaption}
\usepackage{soul}
\usepackage{soul}
\usepackage{hyperref}
{
 \definecolor{BLACK}{gray}{0}
 \definecolor{WHITE}{gray}{1}
 \definecolor{RED}{rgb}{1,0,0}
 \definecolor{GREEN}{rgb}{0,1,0}
\definecolor{dgreen}{rgb}{.1,.6,.1}
\definecolor{BLUE}{rgb}{0,0,1}
 \definecolor{CYAN}{cmyk}{1,0,0,0}
 \definecolor{MAGENTA}{cmyk}{0,1,0,0}
 \definecolor{YELLOW}{cmyk}{0,0,1,0}
 \definecolor{aw}{rgb}{0.2,0.5,0.75}
  }
  \hypersetup{
    colorlinks,%
    citecolor=blue,%
    filecolor=blue,%
    linkcolor=magenta,%
    urlcolor=blue
}
\definecolor{MyGreen}{rgb}{0.0,.5,0.0}

\definecolor{MyDarkRed}{rgb}{0.7,0,0}

\usepackage{mathtools}

\newcommand{\mathsout}[1]% will draw line through middle of #1
{\bgroup\mathchoice
  {\sbox0{$\displaystyle{#1}$}%
    \usebox0\hspace{-\wd0}%
    \rule[0.5\ht0-0.5\dp0-.5pt]{\wd0}{1pt}}%
  {\sbox0{$\textstyle{#1}$}%
    \usebox0\hspace{-\wd0}%
    \rule[0.5\ht0-0.5\dp0-.5pt]{\wd0}{1pt}}%
  {\sbox0{$\scriptstyle{#1}$}%
    \usebox0\hspace{-\wd0}%
    \rule[0.5\ht0-0.5\dp0-.5pt]{\wd0}{1pt}}%
  {\sbox0{$\scriptscriptstyle{#1}$}%
    \usebox0\hspace{-\wd0}%
    \rule[0.5\ht0-0.5\dp0-.5pt]{\wd0}{1pt}}%
\egroup}
\newcommand{\ba}{\begin{eqnarray}}
\newcommand{\ea}{\end{eqnarray}}
\newcommand{\bse}{\begin{subequations}}
\newcommand{\ese}{\end{subequations}}

\newcommand{\B}{{\cal{B}}}

\newcommand{\E}{\mathcal{E}}

\newcommand{\W}{{\cal {W}}}
\newcommand{\FP}{{\mathfrak P}}
\newcommand{\bbq}{\begin{quote}}
\newcommand{\eeq}{\end{quote}}

\newcommand{\CR}{{\cal{R}}}
\newcommand{\CW}{{\cal{W}}}

\newcommand{\EE}{{\cal{E}}}

\newcommand{\GG}{{\cal{G}}}

\newcommand{\HH}{{\cal{H}}}

\newcommand{\A}{{\cal{A}}}
\newcommand{\F}{{\cal{F}}}
\newcommand{\s}{{\cal{S}}}
\newcommand{\Xd}{\mathcal{X}}
\newcommand{\Yd}{\mathcal{Y}}
\newcommand{\Zd}{\mathcal{Z}}
\newcommand{\dd}{{\hbox{d}}}
\newcommand{\One}{\hbox{\tiny{(1)}}}
\newcommand{\Zero}{\hbox{\tiny{(0)}}}

\newcommand{\ToZero}{\!{}^{\hbox{\tiny{(0)}}}}

\newcommand{\sm}{\scriptscriptstyle}
\newcommand{\Pg}{\hat{P}}

\def\t#1#2#3{#1^{\if#2- \else #2 \fi}_{\if#2- \else \:\: \fi #3}}

\newcommand{\ini}{\textrm{\tiny{ini}}}

\def\tdot#1#2#3{\dot{#1}^{\if#2- \else #2 \fi}_{\if#2- \else \:\: \fi #3}}
\def\tddot#1#2#3{\ddot{#1}^{\if#2- \else #2 \fi}_{\if#2- \else \:\: \fi #3}}
\def\ttilde#1#2#3{\tilde{#1}^{\if#2- \else #2 \fi}_{\if#2- \else \:\: \fi #3}}
\def\a#1#2{\eta^{#1}_{\:\:#2}}
\def\ba#1{\boldsymbol{\eta}^{#1}}

\def\adot#1#2{\dot{\eta}^{#1}_{\:\:#2}}
\def\addot#1#2{\ddot{\eta}^{#1}_{\:\:#2}}

\def\two{\mathcal{O}(\Pg^2)}

\font\bigastfont=cmr10 scaled \magstep 2
\def\bdot{\hbox{\bigastfont .}}
\newcommand{\dotaverage}[1]{\left\langle #1 \right\rangle^{\bdot}_\cD}

\newcommand{\cD}{{\cal D}}

\newcommand{\average}[1]{\left\langle #1 \right\rangle_\cD}

\newcommand{\averageHini}[1]{{\left\langle #1 \right\rangle_{\cD_{H\bf i}}}}
\newcommand{\CQ}{{\cal Q}}
\newcommand{\inI}{{\rm I}}
\newcommand{\inII}{{\rm II}}
\newcommand{\inIII}{{\rm III}}

\newcommand{\J}{\mathfrak{J}}

\newcommand{\Joij}{\mbox{}^{(0)\!}\t{\J}{i}{j}}
\newcommand{\JIij}{\mbox{}^{(1)\!}\t{\J}{i}{j}}
\newcommand{\JIIij}{\mbox{}^{(2)\!}\t{\J}{i}{j}}
\newcommand{\JIIIij}{\mbox{}^{(3)\!}\t{\J}{i}{j}}

\newcommand{\Iij}{\mbox{}^{(1)\!}\t{\overline{I}}{i}{j}}
\newcommand{\IIij}{\mbox{}^{(2)\!}\t{\overline{I}}{i}{j}}
\newcommand{\IIIij}{\mbox{}^{(3)\!}\t{\overline{I}}{i}{j}}
\newcommand{\hs}{\hat{\s}}

\newcommand{\z}{z}%{\xi}
\newcommand{\rdv}{\boldsymbol{\xi}}
\newcommand{\zd}{\xi}% dimensionless z
\newcommand{\y}{y}%{\varkappa}
\newcommand{\x}{x}%{\varsigma}
\newcommand{\rv}{\boldsymbol{r}}%{\boldsymbol{\xi}}
\newcommand{\As}{A}%{\xi}

\newcommand{\bW}{\mathfrak{W}}

\begin{document}
\title{Beyond relativistic Lagrangian perturbation theory.\\
I. An exact-solution controlled model for structure formation}
\author{Ismael Delgado Gaspar}
\email{Ismael.DelgadoGaspar@ncbj.gov.pl}
\affiliation{Department of Fundamental Research, National Centre for Nuclear Research, Pasteura 7, 02--093 Warsaw, Poland}
\affiliation{Univ Lyon, Ens de Lyon, Univ Lyon1, CNRS, Centre de Recherche Astrophysique de Lyon UMR5574, F-69007, Lyon, France}
\author{Thomas Buchert}
\email{buchert@ens-lyon.fr}
\affiliation{Univ Lyon, Ens de Lyon, Univ Lyon1, CNRS, Centre de Recherche Astrophysique de Lyon UMR5574, F-69007, Lyon, France}
%%%%%%%%%%%%%%%%%%%%%%%%%%%%%%%%%%%%%%%%%%%%%%%%%%%%%%%%%%%%%%%%%
\author{Jan J. Ostrowski}
\email{Jan.Jakub.Ostrowski@ncbj.gov.pl}
\affiliation{Department of Fundamental Research, National Centre for Nuclear Research, Pasteura 7, 02--093 Warsaw, Poland}
\begin{abstract}
We develop a new nonlinear method to model structure formation in general relativity from a generalization of the relativistic Lagrangian perturbation schemes, controlled by Szekeres (and LTB) exact solutions. The overall approach can be interpreted as the evolution of a deformation field on an inhomogeneous reference model, obeying locally Friedmann-like equations. In the special case of locally one-dimensional deformations, the new model contains the entire Szekeres family of exact solutions. 
As thus formulated, this approach paraphrases the Newtonian and relativistic Zel'dovich approximations, having a large potential for applications in contexts where relativistic degrees of freedom are relevant. 
Numerical simulations are implemented to illustrate the capabilities and accuracy of the model. 
\end{abstract}
%%%%%%%%%%%%%%%%%%%%%%%%%%%%%%%%%%%%%%%%%%%%%%%%%%%%%%%%%%%%%%%%%%
%%%%%%%%%%%%%%%%%%%%%%%%%%%%%%%%%%%%%%%%%%%%%%%%%%%%%%%%%%%%
%04.20.-q	Classical general relativity
%04.20.Jb Exact solutions
%98.80.-k Cosmology
%04.25.Nx  Post-Newtonian approximation; perturbation theory; related approximations
%
\maketitle
%%%%%%%%%%%%%%%%%%%%%%%%%%%%%%%%%%%%%%%%%%%%%%%%%%%%%%%%%%%%%%%
%%%%%%%%%%%%%%%%%%%%%%%%%%%%%%%%%%%%%%%%%%%%%%%%%%%%%%%%%%%%%%%
%%%%%%%%%%%%%%%%%%%%%%%%%%%%%%%%%%%%%%%%%%%%%%%%%%%%%%%%%%%%%%%
\section{Introduction}
\label{Sec:intro}
The problem of cosmological structure formation has been approached by a variety of methods. The exact, inhomogeneous solutions of Einstein's equations provide valuable hints on relativistic effects, absent in Newtonian theory, however, their applications remain limited as the actual cosmic web is much more complex than the scope covered by the exact solutions. Alternatively, the standard cosmological perturbation theory can be used in an attempt to trace the evolution of the density inhomogeneities in the linear regime, the obvious limit being the overdensities or second derivatives of metric perturbations reaching relative values of around unity. These limitations could in principle be overcome with the use of relativistic numerical simulations (see \cite{2020Adameketal} for a recent comparison of performance among the most popular codes currently being developed). There are, however, several drawbacks of these numerical methods with a hard to estimate influence on the outcomes, e.g., global weak field assumption, noncovariance of conformal decomposition, all-time fixed toroidal topology or current computational capacity limitations, to name a few. In light of recently revealed, as well as long known but persistent, tensions haunting modern cosmology, it is very important to improve on the analytical methods providing an  essential counterpart to the numerical efforts. In this spirit, the so-called {\it{silent universe model}} (\cite{silent1}, \cite{silent2}) was proposed and is believed to be able to describe structure formation in the nonlinear regime, from a wide variety of initial data, by an exact method. The {\it{silent universe model}} is based on the specific restriction on the $1+3$ decomposition of Einstein's equations for dust, i.e. the gravitomagnetic part of the projected Weyl tensor is put to zero.  Unfortunately, this model turned out to be insufficient to access the nonlinear regime (because of the absence of rotation and the requirement for the shear to be diagonalizable and have two identical eigenvalues) and, although as desired, it contains several known inhomogeneous exact solutions as subcases, the span of admissible initial data is very restricted. 

In this context, the recent investigation \cite{rza6} consolidated and generalized an earlier insight by Kasai \cite{Kasai1995} on the correspondence between Szekeres class II solutions and the first-order Lagrangian perturbation solutions in relativistic cosmology. The relativistic Lagrangian perturbation schemes have been developed in the series of papers \cite{rza1}, \cite{rza2}, \cite{rza3}, \cite{rza4}, \cite{rza5} and \cite{rza6}. One aspect of this generalization is furnished by the result that an extrapolated version of the first-order solution scheme, in the spirit of the original proposal by Zel'dovich in Newtonian cosmology \cite{ZA}, allows to extend this correspondence to nonlinear functional expressions of the first-order scheme, not only for the density but also for the bilinear metric form, the extrinsic and intrinsic curvatures including their tracefree parts, and other variables. The \textit{exact body} of the functionally extrapolated perturbations is obtained by setting the second and third principal scalar invariants of the deformation field to zero (for details the reader is referred to \cite{rza6}). In the relativistic case, this corresponds to Szekeres class II solutions, while in the Newtonian limit this corresponds to a  class of 3D solutions without symmetry obtained in \cite{buchertgoetz} (with empty background), in \cite{Buchert1989AA} for backgrounds with zero cosmological constant and in \cite{bildhaueretal} including a cosmological constant.

A further insight concerns the way we write the Szekeres class II solutions: 
in the so-called Goode-Wainwright parametrization
\cite{GW1,GW2} this class can be 
written in the form of deviations off a global FLRW 
(Friedmann-Lema\^\i tre-Robertson-Walker) 
background solution. Looking at the spatial average properties of the inhomogeneous deviations, we find admissible initial data for deviations that average out on this background solution. 
This can be realized, as in Newtonian cosmological simulations, by setting periodic boundary conditions on the deviation fields on some scale that is commonly associated with a `scale of homogeneity'. The resulting architecture of such a  (relativistic) simulation has the topological structure of a flat $3$-torus \cite{rza6,NajeraSussman}, very similar to Newtonian simulations, implying integrability of the deformation fields and an on average zero intrinsic scalar curvature (for the notion of integrability of deformations, see \cite{rza4} and \cite{BuchertFocus}).
The results on the correspondence with Szekeres class II solutions carry over to average properties known for constructions with $3$-torus topology, in Newtonian theory and in general relativity \cite{BuchertEhlers,buchert:newton}. This can be summarized by the property of zero cosmological backreaction on the scale set by the size of the toroidal space (for Szekeres class II solutions, see \cite{rza6}). In general, cosmological backreaction can be nonzero \cite{CliftonSussman2019}, and the topology of spatial sections enjoys rich possibilities in general relativity. For constant-curvature models, there are relations to the topology of spatial sections, but in general situations, the scalar curvature function is not tightly constrained; it has a nonzero average and it can also change its sign during the evolution
\cite{buchertcarfora:curvature,GBC}. A different evolution of the scalar curvature compared with the FLRW evolution of the constant curvature also furnishes an explanation of the vividly discussed Hubble tension \cite{hubbletension,Bolejko:2018SilentUniv2} in the standard cosmological model. 

In the present paper, we propose to generalize Lagrangian perturbation schemes. In the example of the first-order scheme we exploit the structure of Szekeres class I solutions that can be written as deviations off a `local background' obeying a Friedmann-type equation. These deviations can still be modeled by the functional expressions developed in relativistic Lagrangian perturbation theory. A `global background' can be defined through a spatial averaging operation of the full solution that leads to nonzero cosmological backreaction, i.e. it allows not only for the impact of the background evolution on the evolution of inhomogeneities as in Newtonian cosmologies, relativistic Lagrangian schemes and standard quasi-Newtonian perturbation theories, but also for the impact of inhomogeneities on the evolution of this global background that is conceived as the average model.
A numerical implementation of this new model will open the door to answering many of the questions raised in the context of inhomogeneous relativistic cosmology, and it may provide a more general architectural setting for relativistic numerical simulations.

We proceed as follows. In Section~\ref{Sec:RZAintro} we recall the Einstein equations in terms of the Lagrangian coframe fields and the definition of the Relativistic Lagrangian Zel'dovich Approximation (RZA). 
Section \ref{SubSec:RelToSzek} is devoted to reviewing the Szekeres models in the Goode and Wainwright parametrization and their relation with RZA. 
In Section \ref{Sec:GRZA}, we develop the proposed generalization of relativistic Lagrangian perturbation schemes to include both classes of the Szekeres solutions. 
Then, its most important subcases, namely the locally one-dimensional solutions, LTB models and RZA, are discussed in Section \ref{Sec:Specialsub}. 
In Section \ref{Sec:ExactbodyNum}, we present a family of simple models and implement numerical simulations aimed at illustrating the capability of the approach to model realistic cosmological structures. 
Finally, in Section \ref{Sec:DiscFinalRemarks}, we put all the elements of our analysis together, discuss their physical motivation and conclude with a summary of the paper and final remarks. 

The text is complemented by five appendices. Appendix \ref{SecApp:MoreOnGWPara} provides additional details about the Goode and Wainwright parametrization. Then, in Appendix \ref{Sec:insightClassI}, 
we have a closer look at Szekeres class I solutions and discuss the relationship between some of their common parametrizations. The model equations are obtained in Appendix \ref{SecApp:GRZAeqns}. 
Appendix \ref{SecApp:EmergingCurvature} contains useful expressions to compute the kinematical backreaction term and its evolution. Finally, in Appendix \ref{App:NumExSol}, we look at the family of models examined in Section \ref{Sec:ExactbodyNum} but from the perspective of exact solutions. 

%%%%%%%%%%%%%%%%%%%%%%%%%%%%%%%%%%%%%%%%%%%%%%%%%%%%%%%%%%%%%%%
%%%%%%%%%%%%%%%%%%%%%%%%%%%%%%%%%%%%%%%%%%%%%%%%%%%%%%%%%%%%%%%
%%%%%%%%%%%%%%%%%%%%%%%%%%%%%%%%%%%%%%%%%%%%%%%%%%%%%%%%%%%%%%%
\section{The relativistic Lagrangian formulation}
\label{Sec:RZAintro}

In the $3+1$ relativistic Lagrangian framework, the Einstein equations' dynamical freedom is completely encoded in the coframe functions $\t{\eta}{a}{i}$ \cite{rza1}.\footnote{Indices $i, j, k, \cdots =1,2,3 \cdots$ denote coordinate indices, while indices $a,b,c \cdots = 1,2,3 \cdots$ are introduced as counters of components, e.g., of vectors or differential forms. In this paper we use units where the gravitational constant and the speed of light are set to $G=c=1$.} Restricting our attention to an irrotational dust source and considering a fluid-flow orthogonal foliation of the spacetime leads to a set of nine purely spatial coframes, in terms of which the line-element reads:
\begin{equation}\label{Eq:gmetricI}
^{(4)} \mathbf{g} = -\mathbf{d}t  \otimes \mathbf{d}t + ^{(3)}\!\mathbf{g} \quad
\hbox{with} \quad ^{(3)}\mathbf{g}=G_{ab} \bm{\eta}^a \otimes \bm{\eta}^b \  .
\end{equation}
Here, the initial metric coefficients are encoded in the Gram's matrix:
\begin{equation}
G_{ij}(\mathbf{X})\equiv g_{ij}(t_{\ini},\mathbf{X}) \ ,
\end{equation}
where $t_{\ini}$ denotes some arbitrary initial time. 

\subsection{The Lagrange-Einstein system}
\label{SubSec:LES}

The Einstein equations are rewritten as a system of $9$ evolution equations and $4$ constraint equations~\cite{rza1,rza4},
\bse\label{Eq:CoframeRelRicci1}
\begin{align}
\label{form_symcoeff}&\hspace{7em} G_{ab} \,\dot{\eta}^a_{[i} \eta^b_{\ j]} = 0 \ ;
 \\
\label{form_eomcoeff}&\frac{1}{2 J} \epsilon_{abc} \epsilon^{ikl}  \left( \dot{\eta}^a_{\ j} \eta^b_{\ k} \eta^c_{\ l} \right)^{\bdot} = -\CR^i_{\ j} + \left( 4 \pi \varrho + \Lambda \right) \delta^i_{\ j} \ ;
 \\
\label{form_hamiltoncoeff}&\frac{1}{2J}\epsilon_{abc} \epsilon^{mjk} \dot{\eta}^a_{\ m} \dot{\eta}^b_{\ j} \eta^c_{\ k} = - \frac{\CR}{2}+ \left( 8\pi \varrho +  \Lambda  \right) \ ;
 \\
\label{form_momcoeff}&\left(\tfrac{1}{J}\epsilon_{abc} \epsilon^{ikl} \dot{\eta}^a_{\ j} \eta^b_{\ k} \eta^c_{\ l} \right)_{||i} = \left(\tfrac{1}{J}\epsilon_{abc} \epsilon^{ikl} {\dot \eta}^a_{\ i} \eta^b_{\ k} \eta^c_{\ l} \right)_{|j} \ .
\end{align}
\ese
We call this system the \textit{Lagrange-Einstein system}. In the equations above, as throughout the text, the overdot represents the partial time derivative (the covariant time-derivative in the flow-orthogonal foliation), while the single and the double vertical slashes stand for the partial and covariant spatial derivatives, respectively. $\CR_{ij}$ denotes the spatial Ricci tensor, $\CR=\t{\CR}{i}{i}$ the spatial scalar curvature, and the determinant of the $3\times3$ coframe matrix is defined as:
\begin{equation}\label{Eq:J-I}
J \equiv \det(\eta^a_{\ i}) \ .
\end{equation}
The exact density field follows from the integration of the continuity equation,
\begin{equation}\label{Eq:denFuncGab}
\varrho=\varrho_{\ini} J^{-1} \ , \quad \hbox{with} \  \quad J=\sqrt{g}/\sqrt{G}\ , 
\end{equation}
and $\varrho_{\ini} =\varrho(t_{\ini})$.\footnote{In this paper we refer to the initial quantities (evaluated at $t_{\ini}$) with the subscript `ini'.}
The expansion tensor can be computed considering its relation to the extrinsic curvature, which in the assumed foliation of the spacetime takes the following form: 
\begin{eqnarray}
\Theta_{ij}&=&-\mathcal{K}_{ij}=\frac{1}{2} \dot{g}_{ij} \ ; 
\\
\label{expansiontensor1}
\t{\Theta}{i}{j}=\t{e}{i}{a} \, \t{\dot{\eta}}{a}{j}\ , &{}& \text{with} \quad \t{e}{i}{a} =\frac{1}{2 J}\epsilon_{a b c}\epsilon^{i k l} \eta^{b}_{\ k} \eta^{c}_{\ l} \ .
\end{eqnarray}
For irrotational dust, the expansion tensor consists of only the expansion scalar (its trace part, $\Theta=\dot{J}/J$) and the shear tensor (its tracefree symmetric part, $\sigma_{ij}$),
\begin{equation}\label{Eq:ThetaShearRel}
\t{\Theta}{i}{j}
=\t{\sigma}{i}{j} \,+\,\frac{1}{3} \Theta \, \t{\delta}{i}{j}
\ .
\end{equation}
The 3D intrinsic curvature and the gravitoelectric and gravitomagnetic parts of the spatially projected Weyl curvature tensor reduce to the following expressions in terms of the coframes~\cite{rza4}:
\begin{subequations}\label{Eq:ExactCofraRelations}
\begin{eqnarray}
-\CR^i_{\ j} &=& \frac{1}{2 J} \epsilon_{abc} \epsilon^{ikl} \left( \dot{\eta}^a_{\ j} \eta^b_{\ k} \eta^c_{\ l} \right)^{\bdot} - \left( 4 \pi \varrho + \Lambda \right) \delta^i_{\ j} \ ; \label{ricci}
 \qquad
\\
- \frac{\CR}{2} &=& \frac{1}{2J}\epsilon_{abc} \epsilon^{mjk} \dot{\eta}^a_{\ m} \dot{\eta}^b_{\ j} \eta^c_{\ k} - \left( 8\pi  \varrho +  \Lambda  \right) \ ; \label{riccitrace}
\\
&&\frac{1}{2J} \epsilon_{abc}\epsilon^{ik\ell} \ddot{\eta}^a_{\ i} \eta^b_{\ k} \eta^c_{\ \ell}   = \Lambda  - 4 \pi {\varrho} \ ;\label{raych}
\\
-  \t{E}{i}{j} & =&  \frac{1}{2J} \t{\epsilon}{-}{abc} \t{\epsilon}{ikl}{} \addot{a}{j} \a{b}{k} \a{c}{l} + \frac{1}{3} \Big(  4 \pi \varrho -  \Lambda \Big) \t{\delta}{i}{j} \ ;  \label{elec}
\\
- \t{H}{i}{j} &=&  \frac{1}{J} \t{G}{-}{ab} \t{\epsilon}{ikl}{} \big( \adot{a}{j\parallel l} \a{b}{k} + \adot{a}{j} \a{b}{k \parallel l} \big) \ . \label{mag}
\end{eqnarray}
\end{subequations}
%
%%%%%%%%%%%%%%%%%%%%%%%%%%%%%%%%%%%%%%%%%%%%%%%%%%%%%%%%%%%%%%%
%%%%%%%%%%%%%%%%%%%%%%%%%%%%%%%%%%%%%%%%%%%%%%%%%%%%%%%%%%%%%%%
%%%%%%%%%%%%%%%%%%%%%%%%%%%%%%%%%%%%%%%%%%%%%%%%%%%%%%%%%%%%%%%
\subsection{Relativistic Zel'dovich Approximation (RZA)}\label{SubSec:RZA}

The relativistic Lagrangian perturbation theory consists of perturbing the trivial coframe set associated with a homogeneous and isotropic spacetime $\t{\eta}{a}{i}=a(t)\t{\delta}{a}{i}$, where $a(t)$ is the solution of the Friedmann equations. Then, RZA emerges as the first-order perturbation of the deformation field, $P^a_{~i}$:
\begin{equation}\label{Eq:CoframesGab}
\bm{\eta}^a=\eta^a_{~i} \mathbf{d}X^i =a(t) \left(\delta^{a}_{~i}+P^a_{~i}\right) \mathbf{d}X^i \ .
\end{equation}
In this approximation, the line-element~\eqref{Eq:gmetricI} takes the following quadratic bilinear form:
\begin{equation}\label{Eq:RZALineElemGab}
g_{ij} = a^2(t) \left[G_{ij} + G_{ab}  \left(\delta^{a}_{~i}  P^{b}_{~j} + \delta^{b}_{~j} P^{a}_{~i}  + P^{a}_{~i} P^{b}_{~j}  \right) \right] \ .
\end{equation}
Consequently, any relevant field (the Ricci and Weyl curvature tensors, rate of expansion, shear, etc.) is functionally evaluated in terms of the coframe fields~\eqref{Eq:CoframesGab}.

%%%%%%%%%%%%%%%%%%%%%%%%%%%%%%%%%%%%%%%%%%%%%%%%%%%%%
%%%%%%%%%%%%%%%%%%%%%%%%%%%%%%%%%%%%%%%%%%%%%%%%%%%%%
%%%%%%%%%%%%%%%%%%%%%%%%%%%%%%%%%%%%%%%%%%%%%%%%%%%%%
\section{Szekeres exact solutions}
\label{SubSec:RelToSzek}
 
The Szekeres models are the most general exact cosmological solution of Einstein's equations \cite{kras1,BolejkoInhCosModels}.\footnote{
This statement requires  some justification.
In the present analysis we follow \cite{BolejkoInhCosModels} and consider as cosmological models those exact solutions of Einstein's equations with at least a nontrivial subclass of FLRW solutions as a limit.
There are other solutions that can be regarded as more general than Szekeres, for example, the Lema\^\i tre model \cite{Lemaitre:1933gd} (although it is spherically symmetric the source is an inhomogeneous perfect fluid) and other models containing heat-flow \cite{Baysal1992Heatflow,Goode1986HeatFlow,LimaMaia1985HeatFlow,NajeraSussman}, viscosity \cite{TomimuraMotta1990aViscosity,MottaTomimura1990bViscosity} or electromagnetic
fields \cite{LimaNobre1989EMfields,TomimuraWaga1987EMfields}. However, in cosmological applications, the field source is usually simplified to a dust fluid and the cosmological constant; then, Szekeres models arise as the most general (exact) cosmological class of solutions.}
These models lack symmetries, albeit their `quasisymmetries' impose significant restrictions on the spacetime's inhomogeneity and anisotropy. The source of the original solution consists of irrotational dust~\cite{Sz75}, which was later generalized by Szafron~\cite{szafron1977inhomogeneous} to include a homogeneous perfect fluid, furnishing the inclusion of the cosmological constant (see \cite{CommentSzeHScaFields} for a discussion of the Szekeres source). For a detailed review of the properties of these solutions see~\cite{kras1,kras2,BKHC2009}. 
The general line-element is commonly expressed as (see \cite{wainwright1977characterization,szafronCollins79,BarnesRowlingson1989SzeInv} for coordinate-independent definitions of the Szekeres solution):
\begin{equation}\label{Eq:GeneralSzeMetric}
\dd s^2=-\dd t^2+e^{2 \alpha} \dd \z^{2} + e^{2 \beta} \left(\dd \x^{2}+\dd \y^{2}\right) \ ,
\end{equation}
where $\alpha=\alpha(t,\rv)$, $\beta=\beta(t,\rv)$, and $\rv=\left(\x,\y,\z\right)$ are comoving coordinates.  
The integration of the Einstein equations is performed by splitting the model into two classes, namely class I, when $\beta_{,\z}\neq0$, and class II, when $\beta_{,\z}=0$. However, once the solution is obtained, the class II can be formulated as a limit of class I~\cite{hellaby1996null}.

In the Goode and Wainwright (GW) parametrization \cite{GW1} the line-element~\eqref{Eq:GeneralSzeMetric} is rewritten as:
\begin{equation}\label{Eq:SzeMetricGW}
\dd s^2=-\dd t^2 + \s^2  \left(\GG^2 \W^2 \dd \z^{2} + e^{2 \nu}  \left( \dd \x^{2} + \dd \y^{2}\right) \right) \ ,
\end{equation}
where the conformal scale factor, $S(t,\z)$, obeys a Friedmann-like equation:
\begin{equation}\label{Eq:FriedmannLikeEqn}
\dot{\s}^2=-k_0 + \frac{2 \mu}{\s} +\frac{\Lambda}{3}\s^2\ .
\end{equation}
Here, $k_0$ is a constant taking the values $0,\pm1$, and $\mu(\z)>0$ is an arbitrary scalar function. Furthermore,
\begin{equation}\label{Eq:GDef}
\GG=\A(\rv)-\F(t,\z)=\A(\rv)-\beta_{+} f_{+}-\beta_{-} f_{-} \ ,
\end{equation}
where $f_\pm$ are the growing and decaying solutions of the following equation:
\begin{equation}\label{Eq:Ftt}
\ddot{\F}+ 2\frac{\dot{\s}}{\s}\dot{\F}-\frac{3\mu}{\s^3}\F=0 \ . 
\end{equation}
The energy-density is given by:
\begin{equation}\label{Eq:Rho2ClassesGenEq}
8 \pi \varrho(t,\rv)=\frac{6 \mu}{\s^3}\left(1+\frac{\F}{\GG}\right) \ ,
\end{equation}
which can be rewritten as:
\begin{equation}
4 \pi \varrho(t,\rv)= \frac{3 \mu \A}{\s^3 \GG} \equiv \frac{M(\rv)}{\s^3 \GG} \ ,
\end{equation}
assigning to the term $3 \mu \A$ the meaning of a conserved rest mass, $M$. 
The functions $f_\pm$ can be regarded as the growing and decaying deviation `modes' at a `fictitious background'
(or reference model)
with initial local density $3\mu=4 \pi \rho$. 

The form of the functions $\A$, $e^{2\nu}$, $\mathcal{W}$, $\beta_{+}(\z)$, $\beta_{-}(\z)$, $f_{+}$ and $f_{-}$ vary depending on the class, and they satisfy:
\begin{description}
\item[ Class I] $\s=\s(t,\z)$, $\mu=\mu(\z)$ and $f_{\pm}=f_{\pm}(t,\z)$.   
\item[ Class II] $\s=\s(t)\equiv a(t)$, $\mu=\text{const.}$ and $f_{\pm}=f_{\pm}(t)$.
\end{description}
For better readability of the text, we display their full expressions in Appendix \ref{SecApp:MoreOnGWPara}. 

The quasispherical branch of the class I solutions has been the most widely used subclass for modeling structure formation, describing nontrivial networks of cosmic structures \cite{Bolejko2006Struformation,IshakNwankwo2008,Bolejko:2009GERG,BolCelerier2010,BolSuss2011,Nwankwo_2011,KrasBol2011,Buckley2013,IshakPRL2013,Vrba:2014,Koksbang2015,Koksbang2015II,sussman2015multiple,sussman2016coarse,Koksbang2017} or even the formation of primordial black holes \cite{HaradaPBHExactSol,gaspar2018black,Coley:2019ylo}. Due to its importance, the physical and mathematical properties of this subclass have also been explored in depth \cite{BonnorTomimura1976evolution,Bonnor1977,HellabyKrasi2002,Hellaby:2007hq,Hellaby:2007hq,WaltersHellaby2012,NonsphericalSzePerts,GIraHellaby2017,Hellaby2017}. The other subclasses have found applications in cosmology and astrophysics as well \cite{ishak2012growth,IshakPeel2012LargeScale,SzeLamddanot0Meures,Meures-Bruni_mnras}, although they have received much less attention. 

%%%%%%%%%%%%%%%%%%%%%%%%%%%%%%%%%%%%%%%%%%%%%%%%%%%%%
%%%%%%%%%%%%%%%%%%%%%%%%%%%%%%%%%%%%%%%%%%%%%%%%%%%%%
%%%%%%%%%%%%%%%%%%%%%%%%%%%%%%%%%%%%%%%%%%%%%%%%%%%%%

\subsection{Szekeres class I models with a normalized scale factor}
\label{SubSec:NormlChi}

In contrast to class II, the class I solution's conformal scale factor is a function of the spatial coordinates, $\s(t_{\ini},\z)=\s_{\ini}(\z)$. To have a normalized initial function in a given Szekeres model, we proceed as follows. 
First, introduce the function $\chi=S^{-1}(t_{\ini},\rv) >0$; then, the conformal scale factor and  the spatial metric  are redefined: 
\begin{equation}
\As(t,\rv) \equiv \s(t,\z) \chi \ ;  
\end{equation}
\begin{equation}
g_{i j}=\As^2~
\mbox{Diag}\left[\left(\A-\F\right)^2\left(\frac{\W} {\chi}\right)^2 ,
\left(\frac{e^\nu}{\chi}\right)^2,
\left(\frac{e^\nu}{\chi}\right)^2
\right] \ , 
\label{Eq:GabNumEx}
\end{equation}
where metric functions are displayed in Appendix \ref{SecApp:MoreOnGWPara}.

In terms of the rescaled scale factor, the Friedmann-like equation 
reads:
\begin{equation}\label{Eq:hatrhoNormA}
\left(\frac{\dot{\As}}{\As}\right)^2=-\frac{\hat{k}(\rv)}{\As^2} + \frac{8 \pi}{3} \frac{  \hat{\varrho}_{b}(\rv)}{\As^3} +\frac{\Lambda}{3}\ , 
\end{equation}
where:   
\begin{equation} \label{Eq:FriedLikeA-1}
\hat{k}(\rv) = k_0 \chi^{2} \ , \quad \frac{4 \pi}{3} \hat{\varrho}_{b}(\rv)= \mu(\z) \chi^3 \ .
\end{equation}
Note that the growing and decaying functions still refer to the local reference background~\eqref{Eq:FriedmannLikeEqn}, and obey the same equation as in~\eqref{Eq:Ftt}:
\begin{equation}\label{eq:fttA}
\ddot{\F}+ 2\frac{\dot{\As}}{\As}\dot{\F}-\frac{4\pi \hat{\varrho}_{b}}{\As^3}\F=0 \ .
\end{equation}
%
%%%%%%%%%%%%%%%%%%%%%%%%%%%%%%%%%%%%%%%%%%%%%%%%%%%%%
%%%%%%%%%%%%%%%%%%%%%%%%%%%%%%%%%%%%%%%%%%%%%%%%%%%%%
%%%%%%%%%%%%%%%%%%%%%%%%%%%%%%%%%%%%%%%%%%%%%%%%%%%%%
\subsection{Relationship between RZA and exact solutions}
\label{SubSec:RZAexact}

In~\cite{rza6} we exploited such a splitting  into `background' and the growing and decaying `deviation modes' to connect the Szekeres class II solution to RZA. The main results of interest for the present article can be summarized as follows:
%
%%%%%%%%%%%%%%%%%%%%%%%%%%%%%%%%%%%%%%%%%%%%%%%%%%%%%%%%%%%%%%%
\begin{itemize}
\item[$(i)$]  RZA contains the class II of the Szekeres exact solutions as a particular case with the identifications:
\begin{subequations}\label{SubEqs:RZASze-classII}
\begin{eqnarray}
\t{P}{a}{i}&=&(-\widetilde{\F}/\widetilde{\A}) \, \t{\delta}{a}{3} \t{\delta}{3}{i} \ ;
\\
G_{ab}&=&\hbox{Diag}\left[  \widetilde{\A}^2, e^{2\nu},e^{2\nu}\right] \ .
\end{eqnarray}
\end{subequations}
In these equations, $\widetilde{\A}\equiv \A-\F_{\ini}$,  $\widetilde{\F}\equiv \F-\F_{\ini}$.
Also, $a(t)$ is the scale factor of the global FLRW background with $4 \pi \varrho_{b}/3= \mu=\mbox{const.}$
%%%%
%%%%
\item[$(ii)$] The class I can be interpreted as a (constrained) superposition of nonintersecting world lines, satisfying RZA's evolution equations. As for the previous class, the associated deformation field and Gram's matrix are given by the identifications: 
\begin{subequations}\label{SubEqs:RZASze-classI}
\begin{eqnarray}
\t{\Pg}{a}{i}&=&(-\widetilde{\F}/\widetilde{\A}) \, \t{\delta}{a}{3} \t{\delta}{3}{i} \ ;
\\
G_{ab}&=&\hbox{Diag}\left[\widetilde{\A}^2 \left(\frac{\W}{\chi}\right)^2, 
\left(\frac{e^\nu}{\chi}\right)^2,
\left(\frac{e^\nu}{\chi}\right)^2
\right]  \ ,
\end{eqnarray}
\end{subequations}
with the proviso that these expressions need to be evaluated for each world line independently. Given an arbitrary world line with fixed comoving coordinates $\rv_i=\left(\x_i,\y_i,\z_i\right)$, $\As(t)\equiv\As(t,\z_i)$ can be regarded as the scale function of an associated FLRW `local background' or `reference model' with initial constant density $\varrho_{b}(t_\ini) = \hat{\varrho}_b(\z_i)$  and  curvature $k_0=\hat{k}(\z)$. Thus, $f_{\pm}(t)\equiv f_{\pm}(t,\z)|_{\z=\z_i}$ are its associated growing and decaying modes.  
\end{itemize}

%%%%%%%%%%%%%%%%%%%%%%%%%%%%%%%%%%%%%%%%%%%%%%%%%%%%%%%%%%%%%%%
%%%%%%%%%%%%%%%%%%%%%%%%%%%%%%%%%%%%%%%%%%%%%%%%%%%%%%%%%%%%%%%
%%%%%%%%%%%%%%%%%%%%%%%%%%%%%%%%%%%%%%%%%%%%%%%%%%%%%%%%%%%%%%%
\section{Generalized Relativistic Zel'dovich Approximation (GRZA)}
\label{Sec:GRZA}

Motivated by the mathematical connection between the Szekeres class II solutions and RZA, we present a new nonperturbative approach that generalizes RZA to contain Szekeres class I. The approach retains the mathematical structure of RZA but without pre-assuming a global background. Instead, we consider a space-dependent conformal scale factor obeying a Friedmann-like equation (as in Eq.~\eqref{Eq:FriedLikeA-1} for Szekeres class I).  

Restricting the analysis to an irrotational dust source, the coframe set~\eqref{Eq:CoframesGab} finds its generalization in
the following expressions:
\begin{subequations}\label{Eq:CoframesGRZA-i}
\begin{eqnarray}
\bm{\eta}^a=\eta^a_{~i} \mathbf{d}X^i =\As \left(\delta^{a}_{~i}+\t{\Pg}{a}{i}\right) \mathbf{d}X^i \ ;
\\
 \As=\As\left(t,\rv \right) \ , \quad  \t{\Pg}{a}{i}=\t{\Pg}{a}{i}\left(t,\rv\right) \ ,
\end{eqnarray}
\end{subequations}
where, as discussed above, $\As$ satisfies a Friedmanian equation for a reference model with curvature and matter density parameters 
$\hat{k}=\hat{k}\left(\rv\right)$ and $\hat{\varrho}_{b}(\rv)$, respectively: 
\begin{subequations}\label{Eq:GRZAFriedLike}
\begin{eqnarray}
2 \ddot{\As}/\As+\dot{\As}^2/\As^2-\Lambda+\hat{k}/\As^2=0 \ ;
\\
\left(\frac{\dot{\As}}{\As}\right)^2=-\frac{\hat{k}}{\As^2} +  \frac{8 \pi }{3} \frac{ \hat{\varrho}_{b}}{\As^3} +\frac{\Lambda}{3}\ .
\end{eqnarray}
\end{subequations}
The line-element keeps the bilinear quadratic mathematical structure for the deformation field, as in Eq.~\eqref{Eq:gmetricI}:
\begin{subequations}\label{Eq:GRZALineElemGab}
\begin{eqnarray}
g_{ij}&=& G_{ab} \eta^{a}_{~i} \eta^{b}_{~j} 
\\
&=& \As^2 \left[G_{ij} + G_{ab}  \left(\delta^{a}_{~i}  \Pg^{b}_{~j} + \delta^{b}_{~j} \Pg^{a}_{~i}  + \Pg^{a}_{~i} \Pg^{b}_{~j}  \right) \right] \ . \qquad
\end{eqnarray}
\end{subequations}
Since at the initial time  $\As$ is normalized and we assume (without loss of generality) that the initial deformation field vanishes ($\Pg^{a}_{~i}(t_\ini)=0$), the Gram's matrix is defined via the initial spatial metric:
\begin{equation}\label{Eq:GabGRZA}
G_{ij}(\rv):=g_{ij}(t_{i},\rv)  \  .
\end{equation}
The subsequent approach consists of {\it(i)} obtaining 
(Lagrange-)linear evolution equations for the 
deformation field 
and, then, {\it(ii)} injecting the formal solution into the exact nonlinear functional expressions.  This scheme retains the original Zel'dovich's extrapolation idea and generalizes RZA to include the whole family of Szekeres models within its locally one-dimensional deformation field limit. In light of this, we call the resulting approach \textit{Generalized Relativistic Zel'dovich Approximation} (GRZA). 

%%%%%%%%%%%%%%%%%%%%%%%%%%%%%%%%%%%%%%%%%%%%%%%%%%%%%%%%%%%%%%%
%%%%%%%%%%%%%%%%%%%%%%%%%%%%%%%%%%%%%%%%%%%%%%%%%%%%%%%%%%%%%%%
%%%%%%%%%%%%%%%%%%%%%%%%%%%%%%%%%%%%%%%%%%%%%%%%%%%%%%%%%%%%%%%
\subsection{Functional evaluation}\label{Sec:FunctEval}

First, let us evaluate all relevant fields as exact functionals of the local deformation and the conformal scale factor. The exact  determinant of the spatial coframe coefficients is given by: 
\begin{eqnarray}\label{Eq:PecDetFunc}
J=\As^3 \J=\As^3\left( \J_0 + \J_1 + \J_2 + \J_3\right) \ ,
\end{eqnarray}
where we have introduced the peculiar-determinant $\J$, and the quantities $\t{\J}{-}{n} \equiv \mbox{}^{(n)\!}\t{\J}{k}{k}$ are defined through:
\begin{subequations}\label{Subeq:DefJij}
\begin{eqnarray}
\Joij&=&  \frac{1}{6} \t{\epsilon}{-}{abc} \t{\epsilon}{ikl}{}  \t{\delta}{a}{j} \t{\delta}{b}{k} \t{\delta}{c}{l} \ ; 
\\
\JIij&=&  \frac{1}{6} \t{\epsilon}{-}{abc} \t{\epsilon}{ikl}{} \t{\Pg}{a}{j} \t{\delta}{b}{k} \t{\delta}{c}{l} 
+\frac{1}{3} \t{\epsilon}{-}{abc} \t{\epsilon}{ikl}{} \t{\delta}{a}{j} \t{\Pg}{b}{k} \t{\delta}{c}{l} \ ;
\\
\JIIij&=& \frac{1}{3} \t{\epsilon}{-}{abc} \t{\epsilon}{ikl}{} \t{\Pg}{a}{j} \t{\Pg}{b}{k} \t{\delta}{c}{l}
+\frac{1}{6} \t{\epsilon}{-}{abc} \t{\epsilon}{ikl}{} \t{\delta}{a}{j} \t{\Pg}{b}{k} \t{\Pg}{c}{l} \ ; 
\qquad
\\
\JIIIij&=&  \frac{1}{6} \t{\epsilon}{-}{abc} \t{\epsilon}{ikl}{} \t{\Pg}{a}{j} \t{\Pg}{b}{k} \t{\Pg}{c}{l} \ .
\end{eqnarray}
\end{subequations}
From its definition, we can see that $\J_0=1$, as expected. Henceforth, the exact inhomogeneous  density field follows from injecting~\eqref{Eq:PecDetFunc} and~\eqref{Subeq:DefJij} into~\eqref{Eq:denFuncGab}.

Next, we express the expansion tensor as a functional of the deformation field. Writing out Eq.~\eqref{expansiontensor1} yields:
\begin{equation}\label{Eq:GRZAExFunExpTen}
\t{\Theta}{i}{j} = 3 \hat{H} \frac{\sum_n \mbox{}^{(n)\!}\t{\J}{i}{j}}{\J} +\frac{1}{\J}\left(\Iij+\IIij+\IIIij\right) \ . 
\end{equation}
To keep the above expression short, we have introduced the shorthand notations:
\begin{subequations}\label{SubEqs:DefOverLinI}
\begin{eqnarray}
\hat{H}&=&\dot \As/\As \ ;
\\
\Iij &=&  \frac{1}{2} \t{\epsilon}{-}{abc} \t{\epsilon}{ikl}{} \t{\dot \Pg}{a}{j} \t{\delta}{b}{k} \t{\delta}{c}{l} \ ; 
\\
\IIij &=& \t{\epsilon}{-}{abc} \t{\epsilon}{ikl}{} \t{\dot \Pg}{a}{j} \t{ \Pg}{b}{k} \t{\delta}{c}{l} \ ;
\\
\IIIij &=& \frac{1}{2} \t{\epsilon}{-}{abc} \t{\epsilon}{ikl}{} \t{\dot \Pg}{a}{j} \t{ \Pg}{b}{k} \t{\Pg}{c}{l} \ .
\end{eqnarray}
\end{subequations}
Taking the trace of the expansion tensor and using a similar notation as before, $\t{\overline{I}}{-}{n} \equiv \mbox{}^{(n)\!}\t{\overline{I}}{k}{k}$, we obtain the functional for the expansion scalar:
\begin{equation}\label{Eq:GRZAExFunExpSc}
\Theta = 3 \hat{H}  +\frac{1}{\J}\left(\overline{I}_1+\overline{I}_2+\overline{I}_3\right) \ . 
\end{equation}
Then, the exact functional for the shear tensor can be computed from~\eqref{Eq:ThetaShearRel},~\eqref{Eq:GRZAExFunExpTen} and~\eqref{Eq:GRZAExFunExpSc}.

The gravitoelectric part of the spatially projected Weyl tensor reads:
\begin{eqnarray}
\t{E}{i}{j} &=& -3\frac{\ddot \As}{\As} \frac{\sum_n \mbox{}^{(n)\!}\t{\J}{i}{j}}{\J}
-2 \hat{H}\frac{\sum_n \mbox{}^{(n)\!}\t{\overline{I}}{i}{j}}{\J} - \frac{\sum_n \mbox{}^{(n)\!}\t{\widehat{I}}{i}{j}}{\J}
\nonumber
\\
&{}& \qquad\qquad\qquad\quad-\frac{1}{3} \left(4\pi \varrho-\Lambda\right)\t{\delta}{i}{j} \ .
\label{Eq:EijExactFun}
\end{eqnarray}
Above, the quantities $\mbox{}^{(n)\!}\t{\widehat{I}}{i}{j}$ are defined as $\mbox{}^{(n)\!}\t{\overline{I}}{i}{j}$ in~\eqref{SubEqs:DefOverLinI}, but replacing $\t{\dot \Pg}{a}{j}$ by $\t{\ddot \Pg}{a}{j}$. The gravitomagnetic part is given by
\begin{eqnarray}
&&\t{H}{i}{j} = -\frac{\t{\epsilon}{ikl}{}}{\As\J} \Bigg\{
\left(\frac{\dot{\As}_{|l}}{\As}+\hat{H}\frac{\As_{|l}}{\As}\right) h_{jk} 
\nonumber
\\
&& \qquad\qquad
+ \hat{H} \left(G_{ja}\t{\Pg}{a}{k||l}+G_{ka}\t{\Pg}{a}{j||l}\right)
\nonumber
\\
&&\qquad\qquad+ 2 \frac{\As_{|l}}{\As}\left(G_{k a}\t{\dot \Pg}{a}{j} +G_{a b}\t{\Pg}{b}{k}\t{\dot \Pg}{a}{j} \right)
+ G_{ka}\t{\dot \Pg}{a}{j||l}
\qquad
\nonumber
\\
&&\qquad\qquad+G_{ab} \left[H \left(\t{\Pg}{a}{j}\t{\Pg}{b}{k}\right)_{||l}+\left(\t{\dot \Pg}{a}{j}\t{\Pg}{b}{k}\right)_{||l}\right]
\Bigg\}
 \ . 
\label{Eq:HWeylFunc}
\end{eqnarray}
In the first line of the above equation, $h_{jk}$ denotes the conformal spatial metric $h_{jk}=g_{ij}/\As^2$, which coincides with the Gram's matrix at the initial time (recall that $A_{\ini}=1$). 

Notice that \eqref{Eq:HWeylFunc} is a complex expression that in general does not identically vanish, except in special cases (like in the Szekeres limit). This suggests a nonsilent character of the GRZA solutions (unless special initial data are imposed) and paves the way for applications to the physics of gravitational waves, see \cite{rza4} for a study of the Lagrangian approach to gravitational waves.

Finally, the functional for the spatial Riemann curvature tensor can be obtained from its well-known expression in terms of partial derivatives of the metric tensor.

%%%%%%%%%%%%%%%%%%%%%%%%%%%%%%%%%%%%%%%%%%%%%%%%%%%%%%%%%%%%%%%
%%%%%%%%%%%%%%%%%%%%%%%%%%%%%%%%%%%%%%%%%%%%%%%%%%%%%%%%%%%%%%%
%%%%%%%%%%%%%%%%%%%%%%%%%%%%%%%%%%%%%%%%%%%%%%%%%%%%%%%%%%%%%%%
\subsection{First-order perturbation scheme}

To determine the first-order scheme, we consider the linearized Lagrange-Einstein system at
a local (Lagrangian) background solution. Linearizing only in the deformation field, i.e., 
neglecting the terms of second- or higher-order in $\t{\Pg}{a}{i}$, we obtain in line with \cite{rza4}:
\begin{subequations}\label{Eq:LinearSystem1}
\begin{eqnarray}
&{}&\left(\frac{\dot{\As}}{\As}\right)^2=-\frac{\hat{k}}{\As^2} + \frac{8 \pi}{3} \frac{\hat{\varrho}_{b}}{\As^3} +\frac{\Lambda}{3}\ ; 
\\
&{}&\qquad \qquad \dot{\Pg}_{[i j]}=0 \ ; \quad
\\
&{}&\t{\ddot{\Pg}}{i}{ j} +3 \hat{H} \t{\dot{\Pg}}{i}{ j} =- \J \left[ \CR \t{{}}{i }{j} -\frac{1}{4}\left( \CR + 2 \frac{\hat{k}}{\As^2}\right) \t{\delta}{i}{ j} \right] \ ;  \quad
\nonumber
\\
\label{SubEq:ddotPij}
\\
&{}&\hat{H} \dot{\Pg} + \frac{4 \pi \hat{\varrho}_{b} }{\As^{3} }\Pg = -\frac{\J}{4} \left(\CR-6\frac{\hat{k}}{\As^2}\right) - \frac{\hat{W}}{\As^{3}} \ ;
\nonumber
\\
\label{SubEq:dotP}
\\
&{}& \frac{\dot{\As}_{|j}}{\As} - \hat{H} \frac{\As_{|j}}{\As} + \frac{1}{\J}\t{\dot{\Pg}}{i}{[i|j]} + \zeta_j = 0 \ ,
\label{Eq:momConstSys1} 
\end{eqnarray}
\end{subequations}
where $\J \equiv J \,A^{-3}$, $\hat{H}$ is the generalized Hubble function, and $\hat{W}$ the trace of the relativistic analog of the Newtonian peculiar-acceleration gradient, defined together with $\zeta_j$ as follows:
\begin{equation}
\hat{W} \equiv -4 \pi \hat{\varrho}_{b} \, \hat{\delta}_\ini \ ,
\;\;
\zeta_j \equiv \frac{1}{2\J}\left(\t{\Gamma}{i}{l j} \t{\dot{\Pg}}{l}{i}-\t{\Gamma}{i}{i l}\t{\dot{\Pg}}{l}{j}\right) \ . 
\label{SubEq:zetaj}
\end{equation}
In the equation above, the Christoffel symbols should be of zeroth-order in the deformation field; otherwise, $\zeta_j$ would be of second or higher order. 
(See Appendix \ref{subsec:connection} for the explicit expression.) 
We have opted for keeping the
peculiar determinant in equations \eqref{Eq:LinearSystem1}-\eqref{SubEq:zetaj}. The importance of this term will be examined elsewhere.
Below we will apply a strict linearization and assume that the curvature terms on the right-hand side of \eqref{SubEq:ddotPij} and \eqref{SubEq:dotP} are small and of the order of $\Pg$.
  
Following the steps of \cite{rza4}, we decompose the deformation field into its  trace, tracefree symmetric, and antisymmetric parts:
\begin{subequations}
\begin{equation}\label{EqApp:DecompPij}
\Pg_{i j}=\Pg_{(i j)} + \Pg_{[i j]} = \frac{1}{3} \Pg G_{ij} +\hat{\Pi}_{i j} + \hat{\FP}_{i j} \ ,
\end{equation}
with the definitions:
\begin{equation}
\hat{\Pi}_{ij} \equiv \Pg_{(ij)} - \frac{1}{3} \Pg  G_{ij} \ ; \quad \hat{\FP}_{ij} \equiv \Pg_{[ij]} \ .
\label{Eq:TraceFreeDecopP}
\end{equation}
\end{subequations}
The linearized system reads:
\begin{subequations}\label{Eq:LinearSystem2}
\begin{eqnarray}
\label{symcondition_order1}
&{}& \dot{\hat{\FP}}_{ij} 
= 0\ ; 
\\
\label{trace_order1}
&{}& \ddot{\Pg} + 3 H \dot{\Pg} = -\frac{1}{4} \left(\CR-6\frac{\hat{k}}{\As^2}\right) \ ; 
\\
\label{sympart_order1} 
&{}&\ddot{\hat{\Pi}}_{ij} + 3 \hat{H} \dot{\hat{\Pi}}_{ij} = - \frac{\tau_{ij}}{\As^2} \ ;
 \\
\label{hamilton_order1} 
&{}& \hat{H} \dot{\Pg} + \frac{4 \pi \hat{\varrho}_{b}}{ \As^{3}} \Pg = -\frac{1}{4} \left(\CR-6\frac{\hat{k}}{\As^2}\right) - \frac{\hat{W}}{\As^{3}}  \ ; \;
\\
\label{momentum_order1} 
&{}& 
\frac{\dot{\As}_{|j}}{\As}- \hat{H} \frac{\As_{|j}}{\As} 
+ \frac{1}{3} \dot{\Pg}_{|j}-\frac{1}{2} \t{\dot{\hat{\Pi}}}{i}{j|i}
 =-\zeta_j \ ,
\end{eqnarray}
\end{subequations}
where $ \tau_{ij}$ denotes the tracefree part of the spatial Ricci curvature,
\begin{equation}
 \tau_{ij} \equiv \CR_{i j} - \frac{1}{3} \CR \, g_{i j}^{\sm{(0)}} \ , \quad \hbox{with} \quad g_{i j}^{\sm{(0)}}=\As^2 G_{ij} \ .
\end{equation}
Keeping only the terms linear in the deformation, Eq.~\eqref{SubEq:zetaj} simplifies to: 
\begin{eqnarray}
\zeta_j&=&
\frac{1}{2}\left(
\frac{A_{| j}}{A} \dot{\Pg}-3 \frac{A_{| l}}{A}\t{\dot{\Pg}}{l}{j} \right) \qquad
\nonumber
\\
&& 
\quad\qquad-\frac{1}{2}\left(\t{\Gamma}{i}{i l}(\mathbf{G})\t{\dot{\Pg}}{l}{j}-\t{\Gamma}{l}{i j}(\mathbf{G}) \t{\dot{\Pg}}{i}{l}\right)
 \ .
 \label{Eq:zeta_j}
\end{eqnarray}
In the above equation, $\t{\Gamma}{i}{j k}(\mathbf{G})$ denotes the Christoffel symbols for the initial spatial metric, see Eq.~\eqref{EqApp:ChristGij}.

Substituting~\eqref{hamilton_order1} into~\eqref{trace_order1}, we obtain the \textit{master equation} for the trace part:
\begin{equation}\label{Eq:MasterEqTracePart}
\ddot{\Pg} + 2 \hat{H} \dot{\Pg} - \frac{4 \pi \hat{\varrho}_{b}}{ \As^{3}} \Pg=  \frac{\hat{W}}{\As^{3}} \ ,
\end{equation}
while the traceless symmetric part obeys~\eqref{sympart_order1}. %
Both coincide in form with their original RZA counterparts (see Eqs. (41) and (24)  of~\cite{rza4}).

%%%%%%%%%%%%%%%%%%%%%%%%%%%%%%%%%%%%%%%%%%%%%%%%%%%%%%%%%%%%%%%
%%%%%%%%%%%%%%%%%%%%%%%%%%%%%%%%%%%%%%%%%%%%%%%%%%%%%%%%%%%%%%%
%%%%%%%%%%%%%%%%%%%%%%%%%%%%%%%%%%%%%%%%%%%%%%%%%%%%%%%%%%%%%%%

\subsection{General initial data setup}
\label{SubSec:initialdata}

To prescribe the initial data, we will proceed as in the standard RZA and identify the time-derivatives of the deformation field with coefficients of the (nonintegrable) generalizations of the Newtonian peculiar-velocity gradient, $\t{\hat{U}}{a}{i}$, and peculiar-acceleration gradient, $\t{\hat{W}}{a}{i}$, respectively:
\begin{subequations}
\begin{eqnarray}
\t{ \Pg}{a}{i}(t_{\ini})&=& 0 \ ;
\\
\t{\dot \Pg}{a}{i}(t_{\ini})&=&\t{\hat{U}}{a}{i} \ ;
\\
\t{\ddot \Pg}{a}{i}(t_{\ini})&=&\t{\hat{W}}{a}{i} -2 \hat{H}_{\ini} \t{\hat{U}}{a}{i} \ .
\end{eqnarray}
\end{subequations}
The initial expansion rate and shear can be computed from the trace of the generalization of the Newtonian peculiar-velocity gradient, $\hat{U}$:
\begin{eqnarray}
\label{Eq:ICsExpansion}
\HH_{\ini}&=&\frac{\Theta_\ini}{3}=\hat{H}_\ini + \frac{\hat{U}}{3} \ ;
\\
\t{\sigma}{i}{j}(t_\ini)&=&\t{\hat{U}}{i}{j} -\frac{\hat{U}}{3}\t{\delta}{i}{j} \ ,
\label{SubEq:Shear}
\end{eqnarray}
where $\hat{H}_{\ini}=\dot{\As}_{\ini}/\As_{\ini}=\dot{\As}_{\ini}$ since $\As_{\ini}=1$.  By construction $\hat{H}_{\ini}$ satisfies a Friedmann-like equation,
\begin{equation}\label{Eq:ICsFriedEqn}
\hat{H}_{\ini}^2=\dot{\As}_{\ini}^2=-\hat{k}(\rv) + \frac{8 \pi}{3} \hat{\varrho}_{b}(\rv) +\frac{\Lambda}{3} \ .
\end{equation}
Here, $\hat{\varrho}_{b}$ is an auxiliary function interpreted as the local background density; then, the physical density can be split as follows:
\begin{equation}\label{Eq:InitialRhoSplit}
4 \pi \varrho_{\ini}= 4 \pi \hat{\varrho}_{b} +4 \pi  \hat{\varrho}_{b} \hat{\delta}_\ini=4 \pi \hat{\varrho}_{b} - \hat{W} \ .
\end{equation}
The initial scalar curvature takes the following form:
\begin{equation}\label{Eq:ICsRicci}
\CR_{\ini} = 6 \hat{k} -4\left(\hat{W} +\hat{H}_{\ini} \hat{U}\right) - \inII\left(\t{\hat{U}}{a}{i}\right)\ ,
\end{equation}
which obeys the exact energy constraint on the initial time slice. The quantity $\inII\left(\t{\hat{U}}{a}{i}\right)$ represents the second principal scalar invariant of the generalized Newtonian peculiar-velocity gradient field. See its definition in Eq.~\eqref{SubEqs:DefInvBij}.

The initial gravitoelectric and -magnetic parts of the (spatially projected) Weyl tensor are given by
\begin{eqnarray}
\t{E}{i}{j}(t_{\ini}) &&=-\t{\hat{W}}{i}{j}+\frac{\hat{W}}{3} \t{\delta}{i}{j}  \ ;
\label{Eq:InitWeylTensor}
\\
\t{H}{i}{j}(t_{\ini}) &&= \!- \t{\epsilon}{ikl}{} 
\left(
\dot{A}_{\ini |l} G_{j k}
+
G_{ka}\t{\hat{U}}{a}{j||l}
\right) \ .
\end{eqnarray}

It should be noticed that  the splitting of the initial density distribution, $\varrho_\ini$, into the associated background density and its contrast is thus far arbitrary in our model, setting an extra degree of freedom in the choice of $\hat{\varrho}_b$. This property is inherited from the Szekeres solution, where $\mu(\z)$ (i.e., $\hat{\varrho}_b$) is an arbitrary free function.

\subsubsection*{Setting the local background density $\hat{\varrho}_b$}  

We choose a physically motivated form of the local background density, namely a constant function with value $\varrho_b (t_{\ini})$, the mass density of an FLRW solution at $t_{\ini}$:
\begin{equation}\label{Eq:LdenBackDef}
\hat{\varrho}_b = \varrho_b (t_{\ini})=\mbox{const.}
\end{equation}
This condition removes the indeterminacy in the local background density and uniquely defines the set of initial conditions. 
The physical motivation behind this choice is the almost-homogeneity that is expected at early cosmic stages. 

Note that Eq.~\eqref{Eq:LdenBackDef} does not imply that $\mu$ is constant in the exact Szekeres limit, which would suppress the growth of structure. 
In fact, $\mu=(4\pi\big/3) \hat{\varrho}_{b} (t_{\ini})  \chi^{-3} \neq \mbox{const.}$ since $\chi(\rv)\neq \mbox{const.}$\footnote{
In the exact solution of class I, $\s_{,\z}=0$ necessarily renders the model an FLRW spacetime.}

%%%%%%%%%%%%%%%%%%%%%%%%%%%%%%%%%%%%%%%%%%%%%%%%%%%%%%%%%%%%%%%
%%%%%%%%%%%%%%%%%%%%%%%%%%%%%%%%%%%%%%%%%%%%%%%%%%%%%%%%%%%%%%%
%%%%%%%%%%%%%%%%%%%%%%%%%%%%%%%%%%%%%%%%%%%%%%%%%%%%%%%%%%%%%%%
\section{Special subcases: locally one-dimensional deformations}
\label{Sec:Specialsub}

So far, we have kept the approach as general as possible. However, we can still introduce some additional assumptions to obtain simpler and more transparent physical models.  
Let us consider the case of a locally one-dimensional deformation field, namely 
\begin{equation}\label{Eq:1-DPzz}
\t{\Pg}{a}{i}(t, \rv )=\Pg(t,\rv)\t{\delta}{a}{\z}\t{\delta}{\z}{i} \ .
\end{equation}
Hence, the linearized evolution equations turn out to be \textit{exact} and the functional expressions introduced in Section~\ref{Sec:FunctEval} become linear,
\begin{subequations}\label{SubEqs:Loc1DFunc}
\begin{eqnarray}
\Theta &=& 3 \hat{H} + \frac{\dot \Pg}{\J} \ ;
\label{Eq:ThetaFunc1DP}
\\
\t{\sigma}{i}{j}&=&\frac{1}{ \J}\left(\t{\dot \Pg}{i}{j} -\frac{\dot \Pg}{3}\t{\delta}{i}{j}\right)
=\Sigma \times \hbox{Diag}\left(1,1,-2\right) \ ;
\label{Eq:ShearFunc1DP}
\\
&{}& \quad\quad\quad\quad\quad\quad\quad\quad\quad  \hbox{with} \quad \Sigma:= -\frac{1}{3}\frac{\dot{\hat{\Pg}}}{\J} \ ;
\\
\CR&=&6 \frac{\hat{k}}{\As^2} -\frac{4}{\J}\left(3\hat{H} \dot \Pg+ \ddot \Pg \right) \ ;
\\
\t{E}{i}{j} &=&-\frac{1}{\J}\left(\t{\ddot \Pg}{i}{j}+ 2 \hat{H} \t{\dot P}{i}{j}\right)
+
\frac{1}{3\J}\left(\ddot \Pg+ 2\hat{H} \dot \Pg\right) \t{\delta}{i}{j} . \quad
\label{Eq:EijLinearP}
\end{eqnarray}
\end{subequations}
The gravitomagnetic part of the Weyl tensor can be obtained by inserting~\eqref{Eq:1-DPzz} into~\eqref{Eq:HWeylFunc}. 

The presence of shear, Ricci, and spatially projected gravitoelectric Weyl tensors with two equal eigenvalues constitutes a characteristic feature of these solutions. 
The importance of the present toy model is two-fold: first, it is the simplest realization of GRZA and its particular case, RZA, and second, it contains the Szekeres models and the subfamily of exact solutions resulting from them.

\subsection{Exact subcases and RZA limit}

The architecture of GRZA allows us to reproduce the class I of Szekeres solutions when the Gram's matrix and the deformation field take the form specified in Eq.~\eqref{SubEqs:RZASze-classI}, 
furnishing a subset of the admissible initial data for the locally one-dimensional models.
Under these assumptions, the constraints are identically satisfied, while Eq.~\eqref{Eq:GRZAFriedLike} and~\eqref{Eq:MasterEqTracePart} are the only nontrivial dynamical equations. Equation~\eqref{Eq:GRZAFriedLike} is nothing more than Eq.~\eqref{Eq:hatrhoNormA}, or~\eqref{Eq:FriedmannLikeEqn}; and \eqref{Eq:MasterEqTracePart} reduces to~\eqref{eq:fttA}, or~\eqref{Eq:Ftt}. 

The well-known LTB models emerge as the spherically symmetric limit of the quasispherical Szekeres models of class I. 
This occurs when the unspecified Szekeres functions in \eqref{SubEqs:RZASze-classI} ($\nu$, $\A$ and $\W$) are set as in~\eqref{SubEq:SzeClasIGW} with $\epsilon=1$ and $c_0, c_1, c_2=\hbox{const.}$ \footnote{The whole class of LTB models is obtained from the Szekeres solution when the dipole vanishes, i.e., when $\Xd=\Yd=\Zd=0$ in Eq.~\eqref{Eq:Szedipole} or equivalently,  $S$, $P$ and $Q$ are constant in Eq.~\eqref{Eq:DipRelSussHellab}.} 

\begin{figure}%[ht]
%\begin{center} 
\includegraphics[width=0.25\textwidth]{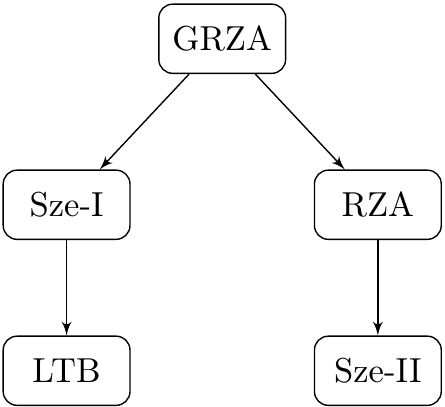}
%\end{center}  
\caption{Special subcases of GRZA. Other exact limits of Szekeres class I (Sze-I) and Szekeres class II (Sze-II) are not represented in the figure; see~\cite{kras1} for their classification. }\label{DiagramGRZA}
\end{figure}

RZA results from GRZA in the limit when the curvature and density parameters of the reference model are constant:
\begin{equation}
\hbox{GRZA} \rightarrow \hbox{RZA}:   
\quad  \hat{k}(\rv) \rightarrow  k_0  
\quad  \land \quad 
\hat{\varrho}_b (\rv) \rightarrow   \varrho_b(t_i)
\ .
\end{equation}
Then, the conformal scale factor is a function of time only, $\As(t,\rv)\rightarrow a(t)$.
In this limit, Szekeres models of class II follow from a simple specialization of the initial data, Eq.~\eqref{SubEqs:RZASze-classII} and \eqref{Eq:SolMetricClassII}.
Figure~\ref{DiagramGRZA} provides a visual representation of the relationship between these solutions; see Fig.~2.1 and 2.4 in~\cite{kras1} for a diagrammatic representation of the exact solutions that can be deduced as a limit of the Szekeres spacetimes.

%%%%%%%%%%%%%%%%%%%%%%%%%%%%%%%%%%%%%%%%%%%%
%%%%%%%%%%%%%%%%%%%%%%%%%%%%%%%%%%%%%%%%%%%% 
%%%%%%%%%%%%%%%%%%%%%%%%%%%%%%%%%%%%%%%%%%%%
%%%%%%%%%%%%%%%%%%%%%%%%%%%%%%%%%%%%%%%%%%%%

\section{A numerical example}
\label{Sec:ExactbodyNum}

To illustrate the potential of GRZA, we explore the numerical solutions of a family of locally 
one-dimensional models, containing the quasispherical Szekeres solution as a particular case.\footnote{A jupyter notebook containing a pedagogical introduction to the numerical example shown in this section is available at 
\url{https://github.com/idgaspar/Beyond_relativistic_Lagrangian_perturbation_theory}.}
A key aspect of these models is that they retain the spatial foliation of the hypersurfaces by nonconcentric 2-spheres, a characteristic feature of the quasispherical Szekeres models, but generalize Szekeres in the way the 2-spheres are arranged. Such a nontrivial foliation leads to more general networks of structures than the standard Szekeres dipolar mass distribution.

\subsubsection{Szekeres-like initial data}

The initial data is specified at the last scattering time. First, we set the free parameters of the set of local Friedmannian reference models. For simplicity, we take for the density field a single global value (while we keep the dependence on the radial coordinate of the curvature parameter) of a conventional FLRW universe at the initial time ($\varrho_{b}(t_\ini):=\varrho_{bi}$):
\begin{equation}
 \hat{\varrho}_{b}(\rv)=\varrho_{bi} \ .
\end{equation}

To work with dimensionless quantities, we take the Hubble parameter at the initial time ($H_\ast=H(t_{ls})$) as the 
timescale and $\ell=2\times 10^{-2}$ Mpc as the characteristic length of the inhomogeneous region. Next, we introduce the following parameters and functions,\footnote{With a slight abuse of notation, in this section we let $\mu$ denote the dimensionless density.}
\begin{equation}\label{Eq:NumExNoDimPar}
\tau=H_{\ast} t \ , \quad \xi=\frac{z}{\ell} \ , \quad  
\mu= \frac{4\pi}{3}\frac{\varrho}{H_{\ast}^2} \ , \quad \varkappa=\frac{\hat{k}}{H_{\ast}^2} \ .
\end{equation}
In these variables, $\mu_{bi}$ denotes the density value of the equivalent FLRW model at the initial time, and the remaining arbitrary free function of the Friedmann-like reference model is given by
\begin{eqnarray}
\hat{\varkappa}(\zd)&=&\hat{\varkappa}_a(\zd) \zd^2 \ ,
\label{Eq:hatk}
\\
\hat{\varkappa}_a(\zd)&=&\hat{\varkappa}_1+\hat{\varkappa}_2 \left(1+\tanh 4\left(\zd-1/2\right) \right) \ ,
\end{eqnarray}
with $\hat{\varkappa}_1=10^{-5}$ and $\hat{\varkappa}_2=37\times 10^{-5}$. For our analysis, we consider these functions in the interval $0 \le\zd\le 1$, $-\infty< x,y <+\infty$. In this model, $\xi$ should be interpreted as a radial comoving coordinate, while $x$ and $y$ are the angular coordinates; see Section \ref{SubSec:PhyIntNumEx} and Appendix \ref{App:NumExSol} below.

The Gram's matrix (initial metric) is taken as
\begin{equation}
G_{ij}=
\mbox{Diag}\left[\left(\A-\hat{\F}_{\ini}\right)^2\left(\frac{\W} {\chi}\right)^2 ,
\zd^2 \E^{-2} ,
\zd^2 \E^{-2}
\right]  \ ,
\end{equation}
where $\chi=\sqrt{\hat{\varkappa}}$ and the metric functions $\E$, $\A$ and $\W$ are chosen as in Szekeres models, Eq.~\eqref{Eq:c012}-\eqref{Eq:c3}, with $S(\zd)=1$, $Q(\zd)=0$ and $P(\zd)$ is a third order polynomial function satisfying $P(0)=P^\prime(0)=P^\prime(\zd_1)=0$ and $P(\zd_1)=-0.6$. Moreover,  $\hat{\F}_{\ini}$ is the initial value of the function 
\begin{equation}\label{Eq:FapproxID}
\hat{\F}=\F+\delta \F \ ,\quad \delta \F=\delta \beta_+ f_+ + \delta \beta_- f_-  \ .
\end{equation}
Here, $\F=\beta_+ f_+ + \beta_- f_-$ is the Szekeres function introduced in Section \ref{SubSec:RelToSzek}, and  
 $\delta\beta_\pm(\rdv)$ are arbitrary functions of the spatial coordinates, $\rdv=\left(\zd,x,y\right)$. In this simulation, we take
\begin{equation}
\delta \beta_+(\rdv)=\alpha \sin^3(\gamma \pi \y) \sin^4(\pi \zd) \ ,
\end{equation}
with $\gamma=3$, and vary $\alpha$. Also, in order to have a model where the voids compensate for the mass excess in the overdensities (i.e., a mass-compensated array of structures), $\delta\beta_-(\rdv)$ satisfies
\begin{equation}
\delta\beta_-(\rdv)=-\frac{\delta\beta_+ f_+(t_\ini,\zd)}{f_-(t_\ini,\zd)} \ .
\end{equation}
Then, the initial energy density is set to
\begin{equation}
\mu_\ini=\mu_{bi}\left(\frac{\A}{\A-\hat{\F}_\ini}\right) \ ,
\end{equation}
which determines the generalized Newtonian peculiar-acceleration divergence,
\begin{equation}
\hat{W}=-3 \mu_{bi} \frac{\hat{\F}_\ini}{\A-\hat{\F}_\ini}  \ .
\end{equation}
The whole set is completed by the initial values of the deformation, $P(t_\ini)=0$, and the generalized Newtonian peculiar-velocity divergence,
\begin{equation}
\hat{U}=-\frac{\dot{\hat{\F}}_\ini}{\A-\hat{\F}_\ini} \ .
\end{equation}

\subsubsection{Physical interpretation}
\label{SubSec:PhyIntNumEx}

This family of models corresponds to a solution with a deformation field of the form~\eqref{Eq:1-DPzz} with
\begin{equation}
\hat{P}(t,\rdv)=-\frac{\hat{\F}-\hat{\F}_{\ini}}{\A-\hat{\F}_{\ini}} \ ,
\end{equation}
containing the exact Szekeres solution in the case $\delta \beta_+(\rdv) = 0$. As the amplitude of $\delta \beta_+(\rdv)$ increases, the model deviates from Szekeres. At the same time, its coupling to the growing mode leads to significantly different evolutions, even for small values of $\delta \beta_+(\rdv)$, compared to the exact solutions.
The function $\delta\beta_+(\rdv)$, its derivatives and the Szekeres dipole functions vanish at the boundary of the domain considered in this simulation, $\zd=\zd_1$, indicating that the solution is an exact LTB model at this shell. 
This property allows matching the inhomogeneous region simulated here to an external Szekeres, LTB, FLRW, or Schwarzschild solution.

Writing out the line-element~\eqref{Eq:GRZALineElemGab}, one can check that the metric of the surfaces $\left\{t=\hbox{const.}, \zd=\hbox{const.}\right\}$ coincides with its equivalent in the exact solution, and then it can be transformed into that of a 2-sphere with radius $\As(t,\zd) \chi(\zd)$ (by the transformation~\eqref{SubEqns:StrProje}). The deviation from the exact solution comes from the metric coefficient $g_{\zd\zd}$, which is generalized to resemble the usual approximation where the growing  (decaying) mode is coupled to a nontrivial function of the spatial coordinates. In this way, the models retain the characteristic foliation by nonconcentric 2-spheres of Szekeres; the term $\delta \beta_+ f_+$ leads to more general arrangements of the 2-spheres. This feature will be examined in detail in Section \ref{SubSec:MappingAndShX}. 

\subsubsection{Estimation of the accuracy}

The GRZA model, although containing exact solutions as special subcases, is in general an approximation. 
The Hamiltonian constraint, or more specifically its violation, can be used to estimate the error generated by this approach. Taking into account that the fluid flow $4-$velocity is the vector normal to our spatial hypersurfaces, we define
\begin{equation}\label{Eq:HamEst}
 \mathbb{H}:= \left(G_{\mu\nu}-8\pi T_{\mu\nu}\right) u^\mu u^\nu=  G_{00}-8\pi T_{00}   \ .
\end{equation}
Then, the function $\mathbb{H}$ vanishes when the solution is exact, and it is nontrivial for approximate models.  
In general, we can tolerate small violations of the constraints as long as they remain below a certain threshold value. 
Nonnegligible values of $\mathbb{H}(\rdv,t)$ can be attributed to either an increase of the numerical error or to the natural failure of the approximate approach as the solution goes deep into the nonlinear regime. 

We exploit the relationship between the present models and the exact solutions to work out a closed-form expression for $\mathbb{H}$. The final result is Eq.~\eqref{Eq:HamError}, displayed in Appendix \ref{App:NumExSol} (where we rewrite this section in the language of exact solutions). Note that the error does not grow linearly in time but exhibits a nontrivial dependence on the `perturbation', reinforcing the nonlinear nature of the current approach.

We note in this context that a self-consistency test, proposed by Doroshkevich \textit{et al.} for the Newtonian Zel'dovich approximation in \cite{Doroshkevichetal73}, see also \cite{Buchert1989AA}, would return a vanishing error, since the comparison of the density field  (i) as a solution of the field equation and (ii) as a solution of the continuity equation, leads to an exact agreement in the case considered here: this error depends on second and third principal scalar invariants of the generalized deformation gradient that both vanish in the class of models considered here.

\begin{figure*}[ht]
     \centering
     \begin{subfigure}[b]{0.31\textwidth}
         \centering
         \includegraphics[width=\textwidth]{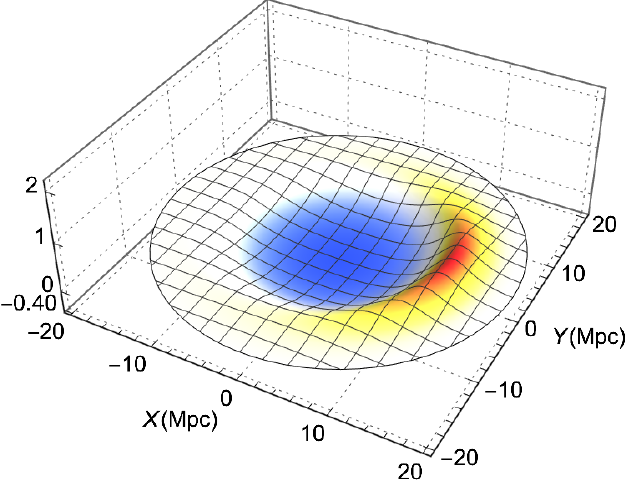}
         \caption{$\alpha=0$}
         \label{fig:DC-alp0Sze}
          \end{subfigure}
       \hfill
       \begin{subfigure}[b]{0.31\textwidth}
         \centering
         \includegraphics[width=\textwidth]{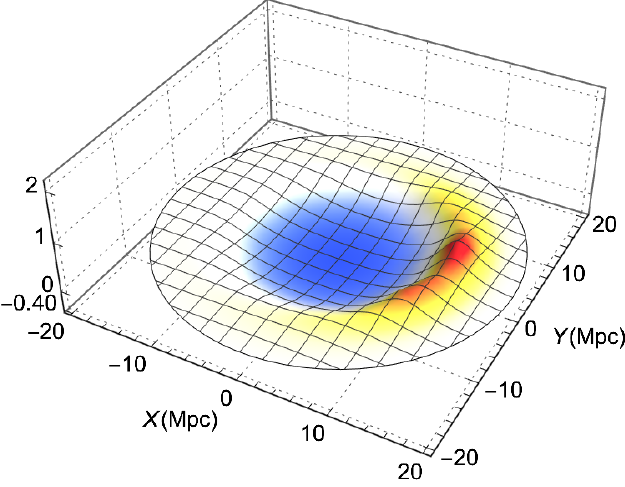}
         \caption{$\alpha=9.6\times 10^{-4}$}
         \label{fig:DC-alp1}
     \end{subfigure}
      \hfill
    \begin{subfigure}[b]{0.31\textwidth}
         \centering
         \includegraphics[width=\textwidth]{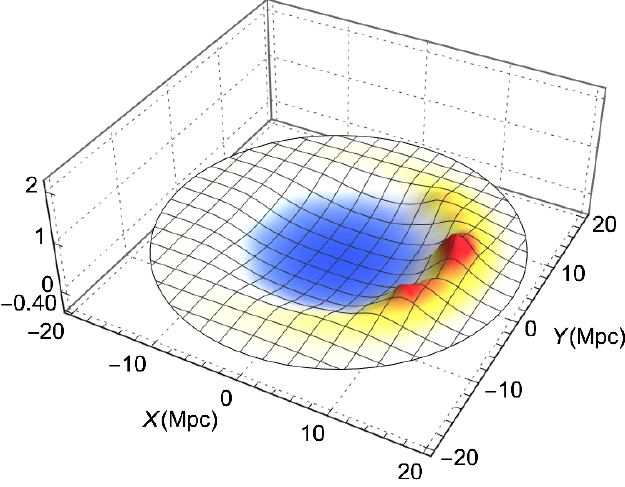}
         \caption{$\alpha=1.9\times 10^{-3}$}
         \label{fig:DC-alp2}
     \end{subfigure} 
    \begin{subfigure}[b]{0.31\textwidth}
         \centering
         \includegraphics[width=\textwidth]{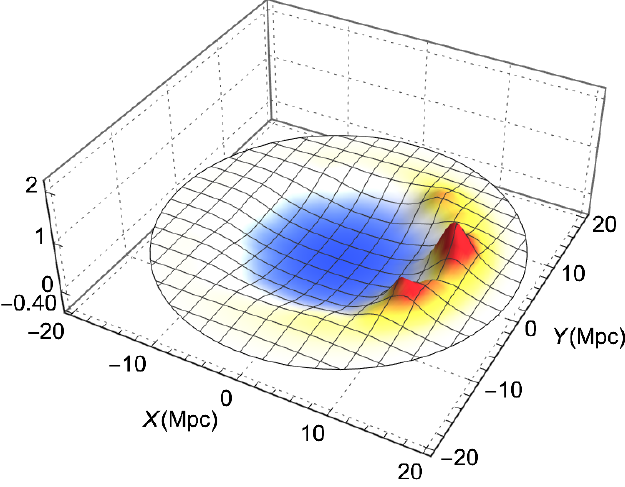}
         \caption{$\alpha=2.9\times 10^{-3}$}
         \label{fig:DC-alp3}
     \end{subfigure} 
    \begin{subfigure}[b]{0.31\textwidth}
         \centering
         \includegraphics[width=\textwidth]{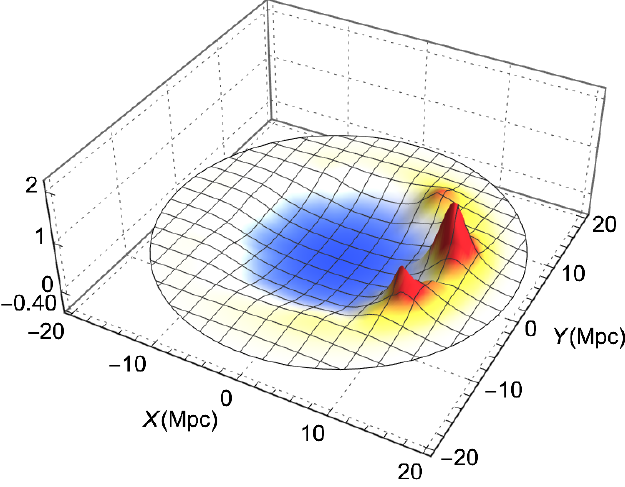}
         \caption{$\alpha=3.8\times 10^{-3}$}
         \label{fig:DC-alp4}
     \end{subfigure} 
    \begin{subfigure}[b]{0.31\textwidth}
         \centering
         \includegraphics[width=\textwidth]{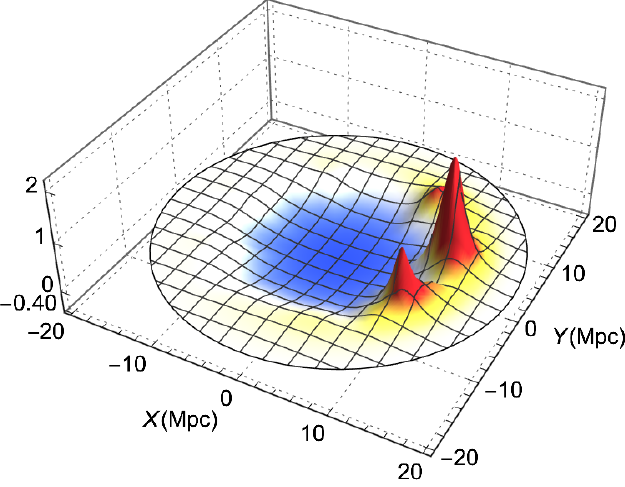}
         \caption{$\alpha=4.8\times 10^{-3}$}
         \label{fig:DC-alp5}
     \end{subfigure} 
     \caption{Equatorial projection of the density contrast $\delta=\varrho/\varrho_b-1$ at the present cosmic time, $t_0$. 
     Coordinates are defined as $X= \As(t_0,\z) \z\cos\phi$ and $Y=\As(t_0,\z) \z \sin\phi$, where $\As(t_0,\z) \z$ is the areal radius. 
     The plots correspond to different values of $\alpha$. For $\alpha=0$, the solution is exact, exhibiting the distinctive dipolar matter distribution of Szekeres models.
     }\label{Fig:snapshots}
\end{figure*}

\begin{figure}[ht]
\includegraphics[width=0.45\textwidth]{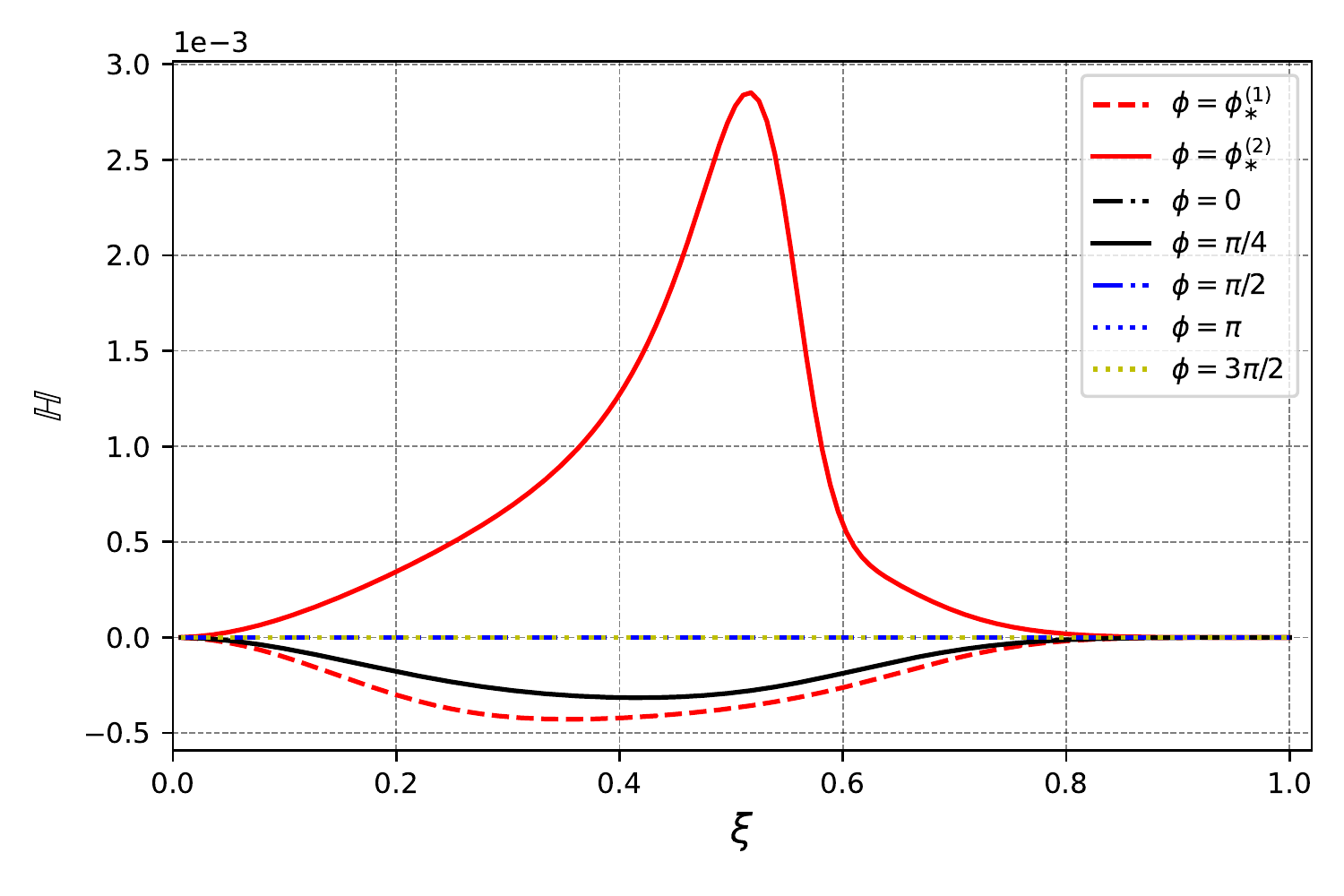}
\caption{\label{fig:errorH}
Estimation of the error from the violation of the Hamiltonian constraint at the present cosmic time. 
Equation~\eqref{Eq:HamEst} is evaluated along the radial rays defined by $\theta=\pi/2$ and $\phi=\left[\phi_{\ast}^{\tiny{\hbox{(1)}}},\phi_{\ast}^{\tiny{\hbox{(2)}}},
0,\pi/4,\pi/2,\pi,3\pi/2\right]$ for the solution with $\alpha=4.8\times10^{-3}$. $\phi_{\ast}^{\tiny{\hbox{(1)}}}$ and $\phi_{\ast}^{\tiny{\hbox{(2)}}}$ denote the azimuthal angles of the two largest overdensities in the order seen by the reader in Fig. \ref{fig:DC-alp5}.
}
\end{figure}

\subsubsection{Numerical results}\label{SubSec:NumRes}

The GRZA model equations are solved in the comoving domain specified above, $\cD_i: 0\le\zd \le 1$.
The inhomogeneous solution is contrasted by a theoretical FLRW model with the standard
cosmological constant and energy density values. Our choice of the initial data ensures that the mass in the initial domain is the same as it would be in an FLRW model ($\varrho_b(t_\ini)=\averageHini{\varrho_{\ini}}$) -- this mass is subsequently conserved through the continuity equation. 
Furthermore, the initial curvature parameter of this FLRW model is taken as the average initial curvature, 
$k_b= \averageHini{\CR_{\ini}}/6$.
In more realistic cosmological simulations, such a background will emerge from the averaging on a sufficiently large compact domain. However, the scales considered here correspond to the size of a typical supercluster, and the background is set by hand. For merely pedagogical reasons, we have chosen to set $\hat{k}$ as a strictly positive function and assumed a background with positive curvature as well.\footnote{
In the GW parametrization, the conformal scale factor has a vertical asymptote at the points where $\hat{k}(\z)=0$. Although the rescaling of the solution removes this singular behavior, we took a
well-behaved model to illustrate how the initial data of Szekeres can be set in Hellaby's formulation, then reparametrized in the language of GW, rescaled, rewritten in the language of GRZA, and finally generalized to obtain the approximate solutions discussed in this section. This is the approach followed in Appendix \ref{App:NumExSol}.}

\begin{figure*}[ht]
     \centering
     \begin{subfigure}[b]{0.305\textwidth}
         \centering
         \includegraphics[width=\textwidth]{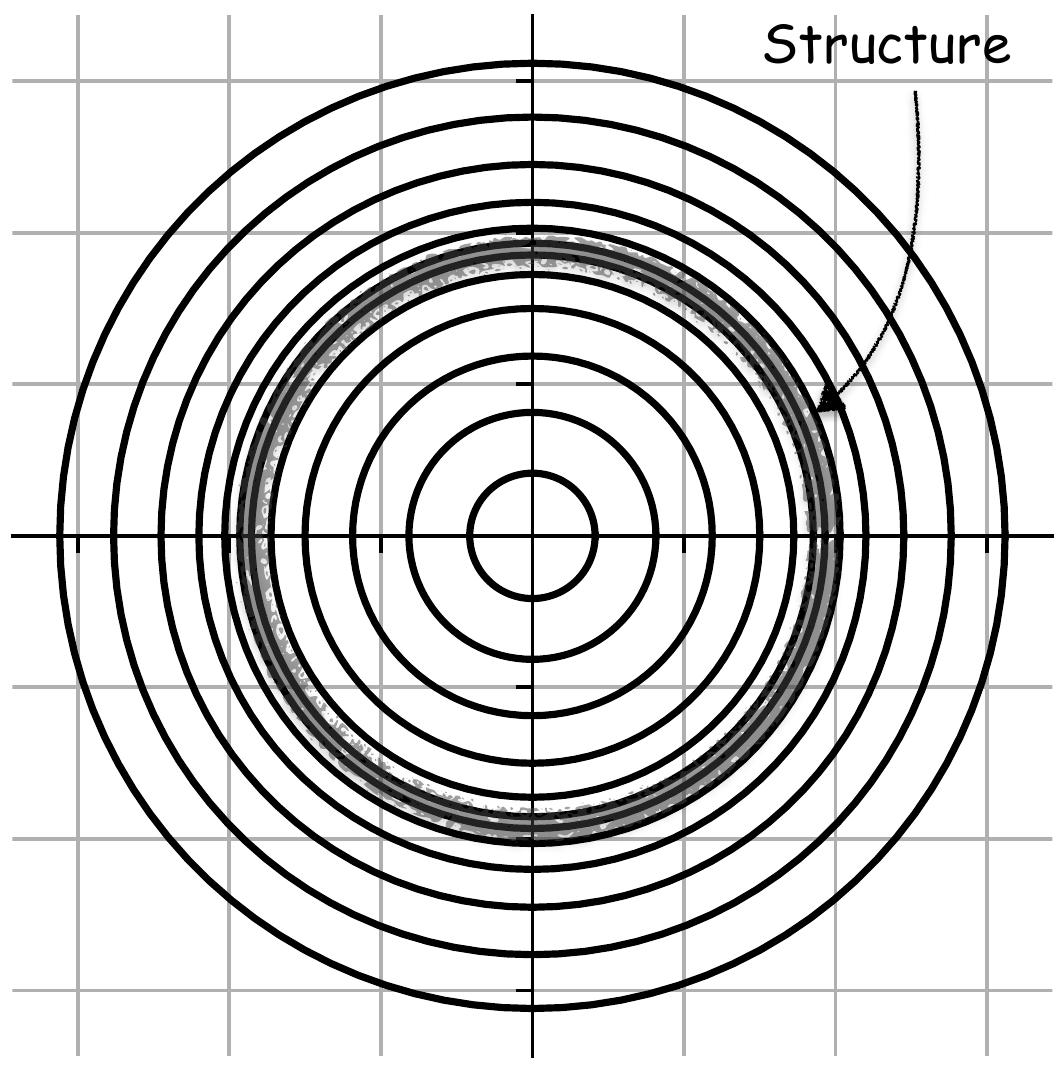}
         \caption{LTB}
         \label{fig:Fol-LTB}
          \end{subfigure}
       \hfill
       \begin{subfigure}[b]{0.29\textwidth}
         \centering
         \includegraphics[width=\textwidth]{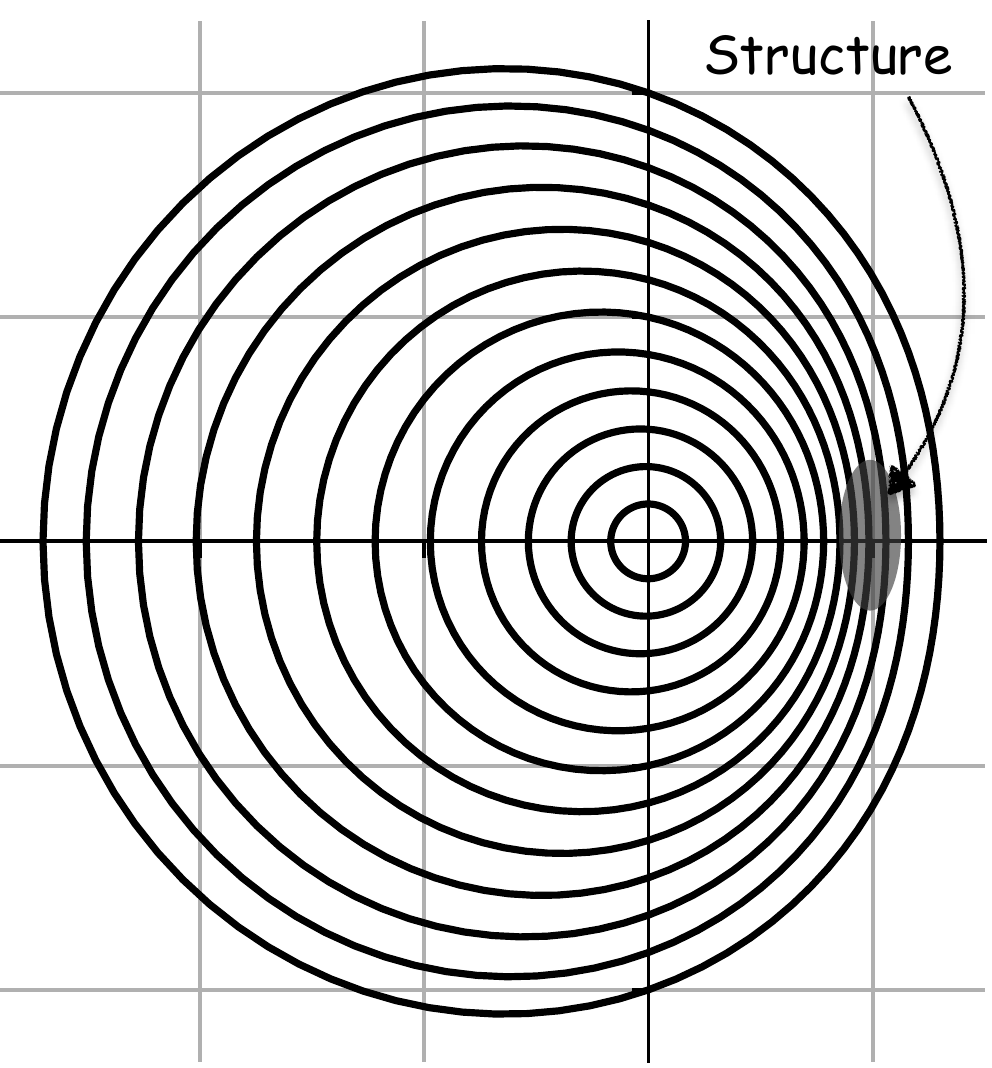}
         \caption{Szekeres}
         \label{fig:Fol-Sze}
     \end{subfigure}
      \hfill
    \begin{subfigure}[b]{0.32\textwidth}
         \centering
         \includegraphics[width=\textwidth]{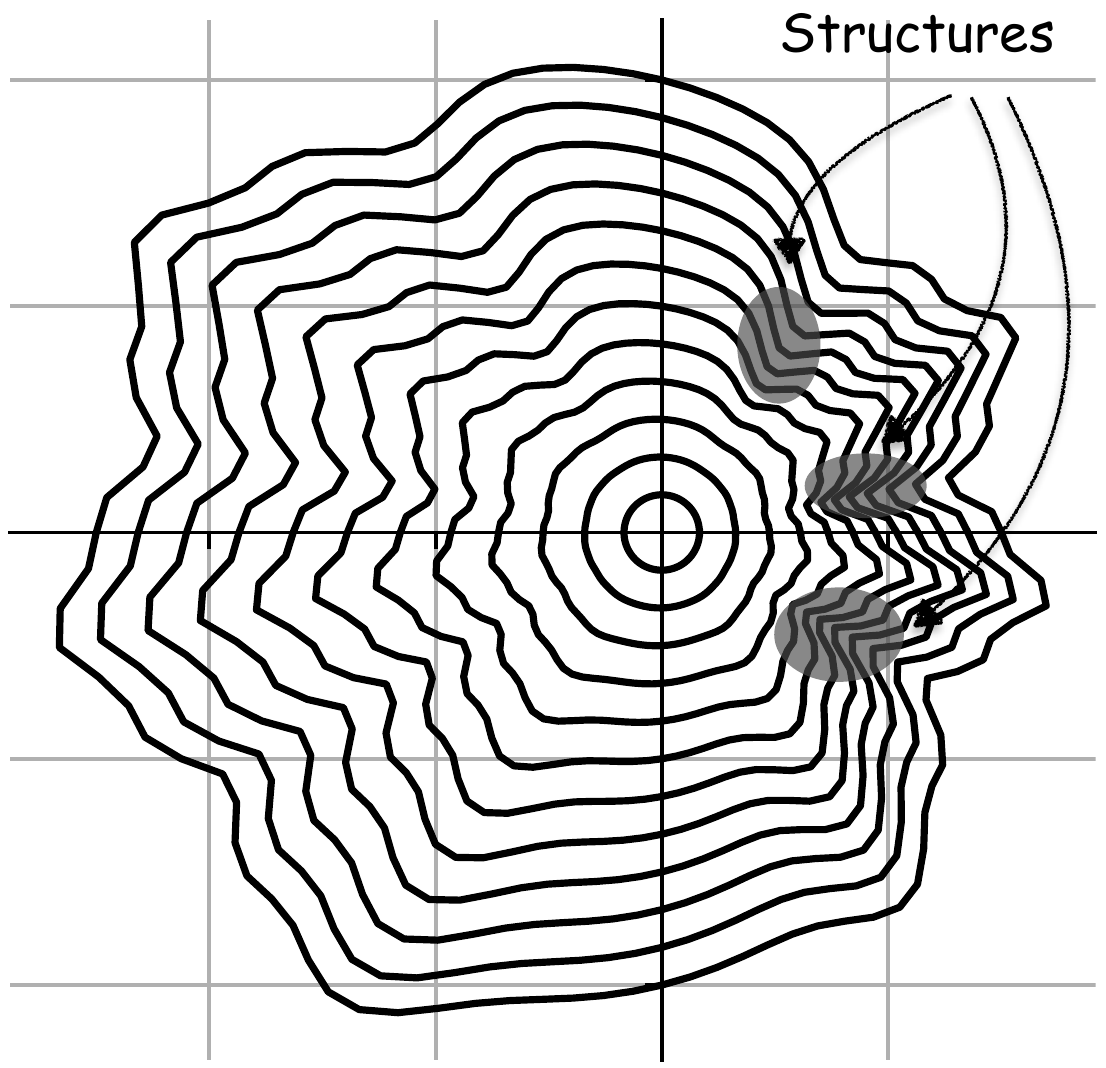}
         \caption{GRZA: Numerical example}
         \label{fig:Fol-GRZA}
     \end{subfigure} 
\caption{Foliation by 2-spheres plotted in terms of the radial proper length. Time slices of LTB and Szekeres solutions and of our numerical example model are leaves of a foliation by topological 2-spheres. Only in the LTB case, the 2-spheres are geometric and concentric. Overdense structures (or density maxima) arise at the loci of points where the proper distance between shells is minimal and becomes zero at shell-crossing. In the Szekeres and more general GRZA cases, these degeneracies also arise along the angular variables but have well-localized spatial positions at points of degeneracy of the deformation field, $\det (\eta^a_{\ i}) =0$. The regions marked in the figure are prone to forming shell-crossings.} 
 \label{Fig:foliation}
\end{figure*}

Figure \ref{Fig:snapshots} shows snapshots of the density contrast at the present cosmic time for the different values of $\alpha$, including $\alpha=0$, the exact Szekeres case. 
In all of these simulations, the error remains small. Figure \ref{fig:errorH} shows $\mathbb{H}$ as a function of $\z$ for different values of the polar angle for the case $\alpha=4.5\times 10^{-3}$. While for the other cases the error remains smaller, being negligible (purely numerical) in the exact case ($\alpha=0$). 

%%%%%%%%%%%%%%%%%%%%%%%%%%%%%%%%%%%%%%%%%%%%%%%%%%%%%%%%%%%%%%%
%%%%%%%%%%%%%%%%%%%%%%%%%%%%%%%%%%%%%%%%%%%%%%%%%%%%%%%%%%%%%%%
%%%%%%%%%%%%%%%%%%%%%%%%%%%%%%%%%%%%%%%%%%%%%%%%%%%%%%%%%%%%%%%

\subsection{Overdense structures and shell-crossings}
\label{SubSec:MappingAndShX}

In the presented models, as in other dust-sourced models, structures emerge from the inhomogeneous expansion and shear of the cosmic fluid. Portions of the fluid expanding faster than the global average distribution with subdominant shear become voids (underdensities), whereas a collapse, i.e. an expansion slower than the global average distribution or a dominant shear leads to clustering (overdensities)   \cite{BolejkoInhCosModels}. In the Szekeres solutions their complex spatial matter distribution can be traced back to overdense spatial patterns in angular directions of a foliation of the constant time slices by nonconcentric 2-spheres, see Figure~\ref{Fig:foliation}.

To gain a better insight into this property let us consider first the LTB models, where the 2-spheres are concentric and structures correspond to spherical dust shells. 
Szekeres models can be thought of as a nonspherical generalization of LTB via the dipole function (see Appendix \ref{SubSec:ModelNetwork}), where the (topological) 2-spheres are no longer orthogonal to the radial rays when they are plotted in terms of proper length \cite{sussman2015multiple}. 
Similarly, the models developed in this section constitute a generalization of Szekeres through the function 
$\delta \beta_+$, furnishing a more general foliation into topological 2-spheres (with ``less quasisymmetries''), and in turn, leading to a more generic network of structures. 
Figure \ref{Fig:foliation}  shows the spatial foliation by 2-spheres in terms of the proper length; we also highlight the regions where the proper distance between adjacent shells is minimal, giving rise to local density maxima, i.e., overdensities. 
Eventually, this distance will go to zero resulting in shell-crossing singularities \cite{kras2}.

It is important to emphasize that the models examined here (and displayed in Fig. \ref{Fig:snapshots})
inevitably lead to shell-crossings, simply because any inhomogeneous physical source that would allow to counteract gravity (such as velocity dispersion, pressure or vorticity) is absent in the basic system of equations \eqref{Eq:CoframeRelRicci1} studied. Although these shell-crossing singularities will emerge in the future, i.e., $t>t_0$, 
their spatial locations coincide with regions where the three overdensities in our example model are located.
Once singularities in the density field emerge, the simulation would ask for regularization of these singularities. We remark here that, in any realization of dust models and their shell-crossings \textit{at finite resolution}, these singularities are smoothed out, so that the distribution of caustics in the density field is indicative for the network of high-density structures, see the methods of realization and illustrations at high resolution in Newtonian models \cite{Buchert1989AA,buchertbartelmann,highresolution}, and for further discussions on shell-crossing singularities in the literature and their morphological classification in generic Lagrangian-to-Eulerian maps, see [sect.VB]\cite{rza1}, as well as the recent review \cite{LagTheorUniver} for related discussions and a list of recent work on shell-crossing singularities in Newtonian models.\footnote{Most of the literature on shell-crossing singularities draws conclusions from Newtonian models. While indicative for shell-crossings in the general-relativistic context, care has to be taken as to whether these studies are applicable. Singularities in the density field have deeper implications in general relativity, since these will in general come with singularities in the geometry, see the discussion in [sect.VB]\cite{rza1}. Questions of extendability of solutions beyond caustics are addressed in e.g. \cite{joshi,nolan}.}

In the mathematical literature, the singularity morphologies are classified topologically. The geometry of caustics emanates from local germs of classified stable caustic patterns, while their global geometry depends on the large-scale distribution of the density field. Locally, the degeneracies of directional scale factors can be used to locate these singularities in the density field (corresponding to degeneracies of the determinant of the coframe deformation field, $\det (\eta^a_{\ i}) = 0$, \textit{cf.} Eq.~\eqref{Eq:denFuncGab}). They can be identified as the eigendirections of the expansion tensor \cite{GW1}: $\dot{\l}_i/\l_i=\Theta_{ii}$ (no summation implied). Then, for the models examined in this section we simply have:
\begin{equation}
\l_3= \As (1+P) \ , \qquad \l_1=\l_2=\As \ .
\end{equation}
In general, shell-crossings arise due to the vanishing of the term $1+P$ (i.e., $\l_3\to 0$) while $\l_1=\l_2>\l_3>0$; resulting in what in cosmology is called a `pancake singularity', i.e. a 2-surface of infinite density in the case of a 3-space. 

%%%%%%%%%%%%%%%%%%%%%%%%%%%%%%%%%%%%%%%%%%%%%%%%%%%%%%%%%%%%%%%
%%%%%%%%%%%%%%%%%%%%%%%%%%%%%%%%%%%%%%%%%%%%%%%%%%%%%%%%%%%%%%%
%%%%%%%%%%%%%%%%%%%%%%%%%%%%%%%%%%%%%%%%%%%%%%%%%%%%%%%%%%%%%%%
\section{General discussion and concluding remarks}
\label{Sec:DiscFinalRemarks}

This section brings together the steps and procedure followed to develop GRZA and discusses their physical interpretation. Our analysis is put in perspective with well-established features of exact solutions and the Lagrangian perturbation theory. 
Finally, we summarize the main points of this article and provide final remarks.

\subsection{Physical motivation}

The relativistic Lagrangian perturbation schemes and exact solutions of Einstein's equations provide complementary approaches to analytically model the large-scale structure formation in the Universe. On the one hand, the Lagrangian description is known for its high accuracy; its first-order solution (known as the Zel'dovich approximation in Newtonian theory for initial data where peculiar-velocity and peculiar-acceleration are assumed to be parallel, so-called `slaving') remains valid beyond the linear regime \cite{LagTheorUniver}. On the other hand, exact solutions are a fundamental tool to understand inhomogeneous cosmology, describing the structure formation in idealized scenarios with symmetries or 
quasisymmetries but otherwise not tied to any particular regime \cite{kras1,kras2}. 

Motivated by our recent results on the correspondence between the relativistic Lagrangian theory of structure formation and the Szekeres family of exact solutions (the first-order Lagrangian perturbation solutions (RZA) contain the entire class II of Szekeres solutions) \cite{rza6}, in this paper, we proposed a generalization of the Lagrangian theory, controlled by Szekeres models of class I.  
This furnishes a new method for structure formation in relativistic cosmology: the generalized Relativistic Zel'dovich Approximation (GRZA).

The GRZA architecture relies on a combination 
of standard elements from the Lagrangian theory and exact solutions. Firstly, we retain the essence of RZA and consider the spatial coframe coefficients as the only variables of the $3+1$ Lagrange-Einstein system \eqref{Eq:CoframeRelRicci1} for irrotational dust in the comoving frame \cite{rza1}. Next, the coframes are split into the expansion of a trivial background model and a deformation field. In this splitting, we introduce a space-dependent scale factor, which obeys a Friedmann-like evolution equation with in general space-dependent density and curvature parameters, Eq.~\eqref{Eq:CoframesGRZA-i}-\eqref{Eq:GRZAFriedLike}. Still, the overall model consists of a deformation field evolving on this inhomogeneous Friedmann-like reference model. As in RZA, the evolution equations are linearized for the deformation, Eq.~\eqref{Eq:LinearSystem1} (or~\eqref{Eq:LinearSystem2}), while the functional evaluation of the physical quantities is kept exact, Section \ref{Sec:FunctEval}. 

The inclusion of such an inhomogeneous Friedmann-like reference model is motivated by the structure of Szekeres class I solutions. However, it can be related to a characteristic feature of the silent solutions of Einstein's equations, where some Bianchi and local rotational symmetric solutions, LTB and Szekeres models are governed by a common set of evolution equations. Their discrimination is reflected in the constraint equations \cite{Bolejko:2018SilentUniv1}. In this way, our space-dependent scale factor allows for fulfilling the constraints associated with more general solutions, including the Szekeres class I exact case. 

To put it differently, GRZA is obtained as a generalization of RZA, where the constant curvature and density of the Friedmann background are extended to be space-dependent functions (leading to a space-dependent scale factor):
$k_0 \rightarrow  \hat{k}(\rv)$ 
and $\varrho_b(t_i) \rightarrow  \hat{\varrho} (\rv)$.
This procedure constitutes the most direct extension of RZA that allows reproducing the entire Szekeres class as an exact limit. 
When the solution is not exact we can control its overall error by estimating the violation of the constraints.

%%%%%
%%%%%%%%%%%%%%%%%%%%%%%%%%%%%%%%%%%%%%%%%%%%%%%%%%%%%%%%%%%%%%%%%%%%

\subsection{Backreaction in cosmological models} 

Let us now turn to the backreaction problem in cosmology, i.e, how small-scale structures can back-react and change the Universe's average expansion rate on large scales. We here refer to the scalar volume-averaging scheme \cite{Buchert2000GERG}, reviewed in \cite{buchert:review, BuchertFocus, buchertrasanen}. As inherited from Szekeres models, GRZA solutions present a nontrivial global backreaction term, which can be computed by inserting~\eqref{Eq:ThetaShearRel}, \eqref{Eq:GRZAExFunExpTen} and \eqref{Eq:GRZAExFunExpSc} into~\eqref{Eqs:BackreactionSys}. However, proper physical models should be compatible with the current observational probes, and the backreaction calculated from classes of exact solutions might be sensitive to these constraints. This is the case of the LTB void models used to fit the observations without dark energy, which have been ruled out by the constraints imposed by the kinematic Sunyaev Zel'dovich effect \cite{SussmanVoidsNoBack,Zibin:2011,Zibin:2011ma,Bull:2011wi}.

The situation will be different when considering a generic model: with GRZA we keep control of approximations made since its construction contains the exact subcases while going beyond the restricted space of initial data. Foremost we emphasize the role of higher scalar invariants contained in GRZA but required to strictly vanish in the exact subcases. Especially for a discussion of cosmological backreaction, a nonvanishing second invariant implies the possibility of a sign change in the backreaction functional: a positive backreaction contributes to acceleration. This generic feature of backreaction is otherwise suppressed, but it is a consequence of the volume-dominance of void regions. Expressed in geometric terms, given a large generic simulation that includes large-volume voids, the average scalar curvature obtains negative contributions mimicking a dark energy-like effect. These remarks show that the implementation of GRZA for generic initial data is key to the future development of the proposed simulations. They will help to quantify cosmological backreaction globally. 

In this context, we think of two questions to be answered. First, does a generic GRZA simulation change the conclusions drawn from LTB models when compared with observational constraints, i.e. what are the quantitative changes by including more degrees of freedom? Second, how backreaction or the emergence of an average negative curvature can solve the tension in the estimation of $H_0$, given the controlled quantification proposed in this paper. How does this compare with silent universe simulations \cite{Bolejko:2018SilentUniv1,
Bolejko:2018SilentUniv2,BolejkoOstrowski2019SilentUniv3}?

These and further questions will be addressed in a forthcoming paper.

%%%%%%%%%%%%%%%
%%%%%%%%%%%%%%%%%%%%%%%%%%%%%%%%%%%%%%%%%%%%%%%%%%%%%%%%%%%%%%%%%%%%%%%%%%%%%%

\subsection{Source of the gravitational field}

The present study has been restricted to models sourced by an \textit{irrotational dust fluid}, dictated by the basic system of equations \eqref{Eq:CoframeRelRicci1}, and aiming at applications in the theory of large-scale structure formation in relativistic cosmology. Both, RZA and GRZA schemes can be extended to include more general sources (isotropic and anisotropic) pressure, velocity dispersion and heat-flow terms, and further sources in the energy-momentum tensor. In these more general cases of a single-fluid model, other nontrivial components should be included in the metric \eqref{Eq:GRZALineElemGab} and a more general four-velocity should be assumed to assure the compatibility of the overall system of equations for the deformation field, e.g. if a nonzero vorticity of the fluid is to be included. 

In concrete terms, the basic system \eqref{Eq:CoframeRelRicci1} can be generalized by including lapse and shift degrees of freedom, as well as a four-velocity that is tilted with respect to the normals to the spatial hypersurface foliation, or if hypersurfaces do not exist, a threading of spacetime can be implemented by using threading lapse and shift functions, see e.g.~\cite{BMR} for the general foliation and threading formalisms. Writing such a general foliated or threaded geometry of spacetime in terms of coframe fields will determine the added geometrical degrees of freedom through the energy-momentum tensor via Einstein's equations, e.g. a pressure gradient will require a nonconstant lapse function, a rotational flow or a non\-diagonal energy-momentum tensor will require shift functions and a tilted four-velocity.
For the RZA model an extension of the general-relativistic Lagrangian perturbation theory to irrotational perfect fluid models with a barotropic equation of state has been investigated  \cite{rza5} using an inhomogeneous lapse function \cite{buchert:fluid}.

%%%%%%%%%%%%%%%
%%%%%%%%%%%%%%%%%%%%%%%%%%%%%%%%%%%%%%%%%%%%%%%%%%%%%%%%%%%%%%%%%%%%%%%%%%%%%%

\subsection{Summary and concluding remarks}

We have presented a new nonlinear method to model large-scale structure formation in the Universe (GRZA). In its development, we merged elements of the Lagrangian perturbation theory and exact solutions; consequently, GRZA contains Szekeres (then LTB) models and RZA as particular limits. This new approach retains the essence of the Lagrangian perturbation theory: the dynamics is described in terms of a deformation field evolving on a cosmological background. But, it generalizes the global FLRW background to an inhomogeneous Friedmann-like reference model. 

The evolution of the GRZA models is formally governed by the same system of equations as the one defining RZA; in this sense, their difference is reflected by the constraints. This similarity in the evolution allows to translate the approach into the Newtonian language, which can be exploited to examine relativistic corrections to the current N-body numerical simulations.
Through their nonperturbative nature, GRZA models are well-suited to explore the impact of the inhomogeneities on the prediction of $H_0$ or the Universe' large-scale average evolution (i.e., to address the Hubble tension and the backreaction problem), to mention a few examples.

To illustrate the method's potential, we provided a family of numerical examples developed as a generalization of the quasispherical Szekeres models. To a certain extent, the path followed in this examination remains pedagogical and explorative. 
More general sets of initial data will open the door to realistic simulations, complementing the current relativistic numerical approaches. We highlight the simplified initial conditions employed to run simulations based on the silent universes approach \cite{bolejko2007evolution,Bolejko:2018SilentUniv1,Bolejko:2018SilentUniv2,BolejkoOstrowski2019SilentUniv3} as a `must-explore' set.

The fact that GRZA contains the entire family of Szekeres (and thus LTB) models is a remarkable theoretical feature. However, this does not necessarily justify its use over RZA, since GRZA has a more complex mathematical structure. Physical applications will ultimately reveal the quality of the approach and whether or not its use is justified. 

\bigskip
%%%%%%%%%%%%%%%%%%%%%%%%%%%%%%%%%%%%%%%%%%%%%%%%%%%%%%%%%%%%%%%
%%%%%%%%%%%%%%%%%%%%%%%%%%%%%%%%%%%%%%%%%%%%%%%%%%%%%%%%%%%%%%%
%%%%%%%%%%%%%%%%%%%%%%%%%%%%%%%%%%%%%%%%%%%%%%%%%%%%%%%%%%%%%%%
%\subsection*
{\bf Acknowledgments:}
This work is part of a project that has received funding from the European Research Council (ERC)
under the European Union's Horizon 2020 research and innovation program (grant agreement ERC advanced grant 740021-ARTHUS, PI: TB). IDG and JJO acknowledge the support of the National Science Centre (NCN, Poland) under the Sonata-15 research grant UMO-2019/35/D/ST9/00342. 
We would like to thank Roberto Sussman for valuable remarks and discussion, and the anonymous referee for valuable suggestions to improve the paper.
IDG and JJO also acknowledge hospitality during their working visits at CRAL-ENS, Lyon. 
% 
%%%%%%%%%%%%%%%%%%%%%%%%%%%%%%%%%%%%%%%%%%%%%%%%%%%%%%%%%%%%%%%
%%%%%%%%%%%%%%%%%%%%%%%%%%%%%%%%%%%%%%%%%%%%%%%%%%%%%%%%%%%%%%%
%%%%%%%%%%%%%%%%%%%%%%%%%%%%%%%%%%%%%%%%%%%%%%%%%%%%%%%%%%%%%%%
%%%%%%%%%%%%%%%%%%%%%%%%%%%%%%%%%%%%%%%%%%%%%%%%%%%%%%%%%%%%%%%
%%%%%%%%%%%%%%%%%%%%%%%%%%%%%%%%%%%%%%%%%%%%%%%%%%%%%%%%%%%%%%%
%%%%%%%%%%%%%%%%%%%%%%%%%%%%%%%%%%%%%%%%%%%%%%%%%%%%%%%%%%%%%%%
%%%%%%%%%%%%%%%%%%%%%%%%%%%%%%%%%%%%%%%%%%%%%%%%%%%%%%%%%%%%%%%
%%%%%%%%%%%%%%%%%%%%%%%%%%%%%%%%%%%%%%%%%%%%%%%%%%%%%%%%%%%%%%%
%%%%%%%%%%%%%%%%%%%%%%%%%%%%%%%%%%%%%%%%%%%%%%%%%%%%%%%%%%%%%%%
%%%%%%%%%%%%%%%%%%%%%%%%%%%%%%%%%%%%%%%%%%%%%%%%%%%%%%%%%%%%%%%
%%%%%%%%%%%%%%%%%%%%%%%%%%%%%%%%%%%%%%%%%%%%%%%%%%%%%%%%%%%%%%%
%%%%%%%%%%%%%%%%%%%%%%%%%%%%%%%%%%%%%%%%%%%%%%%%%%%%%%%%%%%%%%%
%%%%%%%%%%%%%%%%%%%%%%%%%%%%%%%%%%%%%%%%%%%%%%%%%%%%%%%%%%%%%%%
%%%%%%%%%%%%%%%%%%%%%%%%%%%%%%%%%%%%%%%%%%%%%%%%%%%%%%%%%%%%%%%
%%%%%%%%%%%%%%%%%%%%%%%%%%%%%%%%%%%%%%%%%%%%%%%%%%%%%%%%%%%%%%%
%%%%%%%%%%%%%%%%%%%%%%%%%%%%%%%%%%%%%%%%%%%%%%%%%%%%%%%%%%%%%%%
%%%%%%%%%%%%%%%%%%%%%%%%%%%%%%%%%%%%%%%%%%%%%%%%%%%%%%%%%%%%%%%
%%%%%%%%%%%%%%%%%%%%%%%%%%%%%%%%%%%%%%%%%%%%%%%%%%%%%%%%%%%%%%%

\begin{appendix}

%%%%%%%%%%%%%%%%%%%%%%%%%%%%%%%%%%%%%%%%%%%%%%%%%%%%%%%%%%%%%%%
%%%%%%%%%%%%%%%%%%%%%%%%%%%%%%%%%%%%%%%%%%%%%%%%%%%%%%%%%%%%%%%
%%%%%%%%%%%%%%%%%%%%%%%%%%%%%%%%%%%%%%%%%%%%%%%%%%%%%%%%%%%%%%%
\section{The Goode and Wainwright parametrization of Szekeres models}\label{SecApp:MoreOnGWPara}

\noindent
We start by giving the relevant functions for class I ($\beta_{,\z}\neq0$):
\begin{subequations}\label{SubEq:SzeClasIGW}
\begin{eqnarray}
\s=\s(t,\z)\ , \quad  \hbox{with} \quad \s_{,\z}\ne0 \ ,
\\
f_{\pm}=f_{\pm}(t,\z)\ , \quad t_B= t_B (\z)\ , \quad \mu=\mu(\z) \ ,
\end{eqnarray}
and
\begin{eqnarray}
&{}&e^\nu=f(\z)\E^{-1} \ ;
\\
&{}&\E = \Big[c_0(\z)(\x^2+\y^2)+2c_1(\z)\x+2c_2(\z)\y+c_3(\z)\Big]^{-1} ; 
\nonumber
\\
\label{Eq:eToNuGW}
\\
&{}& c_0 c_3-c_1^2-c_2^2=\epsilon/4\ ,\,\quad \quad \epsilon=0\ ,\pm 1 \ ; \label{SubEq:Rel-ci}
\\
&{}& \A=f\nu_{,\z}-k_0\beta_+ \ , \quad \;  \W^2=(\epsilon-k_0 f^2)^{-1} \ ; \; \label{Eq:AandWclassIGW}
\\ 
&{}& \beta_+=-k_0 f \mu_{,\z}/(3\mu)\ ,\, \quad \beta_-=f \, t_{B,\z} \ .
\label{ClassIBeta}
\end{eqnarray}
\end{subequations}
The arbitrary function $ t_B (\z)$  is the inhomogeneous `big bang time'. 

Correspondingly, for class II ($\beta_{,\z}=0$):
\begin{subequations}\label{Eq:SolMetricClassII}
\begin{eqnarray}
\s=\s(t)\ , \quad f_{\pm}=f_{\pm}(t)\ , \quad  t_B, \ \mu=\hbox{const.} \ ;
\\
 k_0=0,\pm1\ ,\quad \W=1 \ ;
 \\
e^\nu=\left[1+\frac{k_0}{4}(\x^2+\y^2)\right]^{-1} \ ,
\label{Eq:etonuClassII}
\end{eqnarray}
and
\begin{equation}\label{Eq:DefAclassII}
  \A = \left\{ 
  \begin{array}{l l}
    e^\nu \Big[ c_0(\z) \left(1-\frac{k_0}{4}(\x^2+\y^2)\right) + c_1(\z)\x \\
    \hspace{.9cm}  +c_2(\z)\y \Big]-k_0\beta_{+}\ ,\; \hbox{for} \; k_0=\pm 1\ ;
    \\
    c_0(\z)+c_1(\z)\x+c_2(\z)\y
    \\
    \hspace{.9cm}-\beta_{+} (\z) (\x^2+\y^2)/2 \ , \;  \hbox{for} \; k_0=0 \ ;
  \end{array} \right.
\end{equation}
\end{subequations}
$c_i(\z)$ and $\beta_{\pm}(\z)$ arbitrary and $\mu$, $ t_B=\text{const.}$

%%%%%%%%%%%%%%%%%%%%%%%%%%%%%%%%%%%%%%%%%%%%%%%%%%%%%
%%%%%%%%%%%%%%%%%%%%%%%%%%%%%%%%%%%%%%%%%%%%%%%%%%%%%
%%%%%%%%%%%%%%%%%%%%%%%%%%%%%%%%%%%%%%%%%%%%%%%%%%%%%
%
\section{A deeper insight into Szekeres models of class I}
\label{Sec:insightClassI}

% %%%%%%%%%%%%%%%%%%%%%%%%%%%%%%%%%%%%%%%%%%%%%%%%%%%%%
% %%%%%%%%%%%%%%%%%%%%%%%%%%%%%%%%%%%%%%%%%%%%%%%%%%%%%
% %%%%%%%%%%%%%%%%%%%%%%%%%%%%%%%%%%%%%%%%%%%%%%%%%%%%%

Szekeres models of class I are traditionally examined in an LTB-like parametrization introduced by Hellaby \cite{hellaby1996null}. In this representation, the line-element is given by 
\begin{equation}
\dd s^2 =-\dd t^2 + \frac{\left(\Phi_{,\z}-\Phi \E_{,\z}/\E\right)^2}{\epsilon-k(\z)} \dd \z^2 +\frac{\Phi^2}{\E^2}\left(\dd \x^2 + \dd \y^2\right) \ ,
\end{equation}
with
\begin{equation}\label{AppEq:ESze}
 \E=\frac{S(\z)}{2}\left(\left(\frac{\x-P(\z)}{S(\z)}\right)^2 + \left(\frac{\y-Q(\z)}{S(\z)}\right)^2+\epsilon\right) \ ,
\end{equation}
where the latter expression is just a reparametrization of \eqref{Eq:eToNuGW}.
Here, $\Phi=\Phi(t,\z)$, $\epsilon=\pm 1, 0$ (for quasispherical, quasihyperbolic and quasiplane models, respectively), and $k(\z)$, $P(\z)$, $Q(\z)$ and $S(\z)$ are arbitrary functions.

The function $\Phi$ obeys a Friedmann-like equation,
\begin{subequations}\label{AppEq:dotPhi}
\begin{equation}
\dot \Phi^2=-k(\z)+\frac{2 M(\z)}{\Phi}+\frac{1}{3}\Lambda \Phi^2 \ ,
\end{equation}
which can be integrated to give 
\begin{equation}\label{AppEq:ForIntPhi}
t-t_{B}(\z)=\int_0^\Phi \frac{\dd \tilde{\Phi}}{\sqrt{-k+2M/\tilde{\Phi}+\frac{1}{3}\Lambda \tilde{\Phi}^2}} \ .
\end{equation}
\end{subequations}
$M(\z)$ and $t_B(\z)$ are arbitrary integration functions, identified as the effective gravitational mass  and the time of the initial singularity (the `big bang time').

In these variables the density reads:
\begin{equation}
4\pi \varrho= \frac{M_{,\z}-3M \E_{,\z}/\E}{\Phi^2\left(\Phi_{,\z}-\Phi\E_{,\z}/\E\right)}  \ ,
\end{equation}
while the expressions for the expansion, shear and curvature tensors can be seen in Section 2.3.2 of Ref. \cite{BKHC2009}.

%%%%%%%%%%%%%%%%%%%%%%%%%%%%%%%%%%%%%%%%%%%%%%%%%%%%%
%%%%%%%%%%%%%%%%%%%%%%%%%%%%%%%%%%%%%%%%%%%%%%%%%%%%%
%%%%%%%%%%%%%%%%%%%%%%%%%%%%%%%%%%%%%%%%%%%%%%%%%%%%%

\subsection{Coordinate system}
\label{SubSec:CoordSys}

The coordinate system of the standard representation of Szekeres models has the physical interpretation of a stereographic projection of a sphere ($\epsilon=1$), plane ($\epsilon=0$), or hyperboloid ($\epsilon=-1$)~\cite{kras1,kras2,Bolejko2006Struformation}.
In the quasispherical case, the transformation
\begin{subequations}\label{SubEqns:StrProje}
\begin{eqnarray} 
\x &=& P+ S \cot \left( \vartheta/2 \right) \cos (\phi) \ ; 
 \\
\y &=& Q + S \cot \left( \vartheta/2 \right) \sin (\phi)  \ ,
 \end{eqnarray}
 \end{subequations}
turns the metric of the surfaces $\left\{t=\mbox{const.},\z=\mbox{const.}\right\}$ into that of a 2-sphere,
\begin{equation}
\frac{\Phi^2}{\EE^2}\left(\dd \x^2+\dd \y^2\right)=\Phi^2 \left(\dd \vartheta^2+\sin^2\vartheta \dd \phi^2 \right) \ .
\end{equation}
%

%%%%%%%%%%%%%%%%%%%%%%%%%%%%%%%%%%%%%%%%%%%%%%%%%%%%%
%%%%%%%%%%%%%%%%%%%%%%%%%%%%%%%%%%%%%%%%%%%%%%%%%%%%%
%%%%%%%%%%%%%%%%%%%%%%%%%%%%%%%%%%%%%%%%%%%%%%%%%%%%%

\subsection{Modeling of the cosmic network}
\label{SubSec:ModelNetwork}

It is often convenient to employ spherical coordinates to examine the dynamics of the quasispherical models. Performing the stereographic transformation of coordinates~\eqref{SubEqns:StrProje} and $\z=r$, one finds that all the information about the departure from spherical symmetry is contained in the `Szekeres dipole' function,
\begin{equation}\label{Eq:Szedipole}
\bW :=\frac{r \EE'}{\EE}= -\Xd\,\sin\vartheta\,\cos\phi- \Yd\,\sin\vartheta\,\sin\phi-\Zd\,\cos\vartheta \ ,\quad
\end{equation}
where we have defined $\Xd$, $\Yd$ and $\Zd$ as
\begin{equation}\label{Eq:DipRelSussHellab}
\Xd= \frac{r P^\prime}{S};\qquad \Yd = \frac{r Q^\prime}{S}; \qquad \Zd=\frac{r S^\prime}{S} \ ,
\end{equation}
which in terms of the GW free functions take the form,
\begin{eqnarray}
\Xd = 2 \z \left(\frac{c_1 c_0^\prime}{c_0}-c_1^\prime \right) \ , \; \Yd &=& 2 \z \left(\frac{c_2 c_0^\prime}{c_0}-c_2^\prime \right) \ ,
\nonumber
\\
  \Zd &=&-\z \frac{ c_0^\prime}{c_0} \ .
\end{eqnarray}
For more details see Ref.~\cite{SussBol2012}, where Sussman and Bolejko introduced a description of this subclass employing 
 weighed proper volume averages of the local scalars (so-called q-scalars) and their fluctuations. 
 The advantages of this notation were exploited in Ref.~\cite{sussman2015multiple} to obtain sufficient conditions for the existence of extrema of the Szekeres scalars. In particular, the functions $\Xd$, $\Yd$ and $\Zd$ (and in turn $S$, $P$ and $Q$) set the angular position of the extrema in the curve defined by $\bW_{,\vartheta}=\bW_{,\phi}=0$:
\begin{equation}  
\B_{\pm}(r )=[r,\,\vartheta_{\pm}(r ),\,\phi_{\pm}( r)], \label{Eq:BpmAngExtre}
\end{equation}
where  $\vartheta_{\pm}$ and $\phi_{\pm}$ denote two antipodal positions for each value of $r$,
\begin{subequations}
\begin{eqnarray}
\phi_{-} &=& \arctan \left(\frac{\Yd}{\Xd}\right),\;\; \vartheta_{-}= \arccos\left(\frac{\Zd}{\sqrt{\Xd^2+\Yd^2+\Zd^2}}\right) \ ,
\nonumber
\\
\label{Eq:AngExtPosm}
\\
 \phi_{+} &=& \pi+\phi_{-},\qquad \vartheta_{+}= \pi-\vartheta_{-}  \ .
 \label{Eq:AngExtPosp}
\end{eqnarray}
\end{subequations}
The extrema  of all the scalars lie along the curve $\B_{+}$.  
And we have overdensities (/voids) in a given interval of $r$, if $M_{,r}>0$ (/$M_{,r}<0$). Section~\ref{SubSec:NumRes} considers $S=1$ and $Q=0$; thus, the extremum of the exact case is located on the equatorial plane and along the x-axis.

The limit $\bW=0$ defines a unique spherically symmetric LTB model, where the 2-spheres foliating the time slices become concentric (see Section \ref{SubSec:MappingAndShX}). This idea was exploited in \cite{sussman2015multiple} to introduce the term  term `LTB seed model'; then, every Szekeres model can be developed from an `LTB seed model' by specifying the dipole parameters.  

%%%%%%%%%%%%%%%%%%%%%%%%%%%%%%%%%%%%%%%%%%%%%%%%%%%%%
%%%%%%%%%%%%%%%%%%%%%%%%%%%%%%%%%%%%%%%%%%%%%%%%%%%%%
%%%%%%%%%%%%%%%%%%%%%%%%%%%%%%%%%%%%%%%%%%%%%%%%%%%%%

\subsection{Relationship between the LTB-like and GW parametrizations}
\label{SubSec:RelHella-GW}

To obtain the GW parametrization, firstly, we introduce the constant $k_0=|k(\z)|/k(\z)$, and define the metric function $f(\z)$ as
\begin{equation}\label{Eq:f2defk}
k_0 f^2(\z)=k(\z) \quad \mbox{when} \quad k_0\neq 0 \ .
 \end{equation}
Otherwise, $f$ is arbitrary. 

The remaining functions relate to those in the LTB-like representation as follows \cite{IshakPeel2012LargeScale}:\footnote{When $k=0$, the scale factor is defined by $\s=\Phi/\mu^{1/3} \implies \mu=1$~\cite{kras1,kras2}.   }
\begin{subequations}\label{AppSEqns:ParamGW}
\begin{eqnarray}
c_0=1/(2 S) \ , \; c_1=-P/(2 S) \ , \;  c_2=-Q/(2 S) \ ; \qquad
\label{Eq:c012}
\\
c_3=(P^2+Q^2+\epsilon\, S^2)/(2 S) \ ; \qquad
\label{Eq:c3}
\\
S=\Phi/f  \ , \;\;  \mu=M/f^3 \ ; \;\; \W=\left(\epsilon-k\right)^{-\frac{1}{2}} \ ; \qquad
\\
\GG = f \left(\frac{\Phi_{,\z}}{\Phi} -\frac{\EE_{,\z}}{\EE}\right) \ .\qquad\;
\label{AppSub:G1}
\end{eqnarray}
\end{subequations}
To complete the analysis, we recall some results of~\cite{rza6}. 
Focusing on the case $k\neq 0$, Eq.~\eqref{AppSub:G1} gives
\begin{eqnarray}
\GG &=& f \left(\frac{\Phi_{,\z}}{\Phi} -\frac{\EE_{,\z}}{\EE}\right) 
=  \left(f\frac{\s_{,\z}}{\s} +f_{,\z}-f\frac{\EE_{,\z}}{\EE}\right)
\nonumber
\\
&=&f \frac{\s_{,\z}}{\s}+f \nu_{,\z} \ ,
\end{eqnarray}
with $e^\nu:=f/\EE$.
Under the parametrization~\eqref{AppSEqns:ParamGW}, Eq.~\eqref{AppEq:dotPhi} transforms into~\eqref{Eq:FriedmannLikeEqn}, which 
has the following formal integral (similar to~\eqref{AppEq:ForIntPhi}):
\begin{subequations}
\begin{equation}\label{AppEq:IntSParm}
t-t_B(\z)=\int^{\s}_0\frac{\dd \hs}{\left(-k_0 + \frac{2 \mu}{\hs} +\frac{\Lambda}{3}\hs^2\right)^{1/2}} \ ,
\end{equation}
where, as before, $t_B(\z)$ is the `big bang time' and
\begin{equation}\label{AppEq:dotSParm}
\dot{\s}=\left(-k_0 + \frac{2 \mu}{\s} +\frac{\Lambda}{3}\s^2\right)^{1/2} \ .
\end{equation}
Differentiating~\eqref{AppEq:IntSParm} (and substituting~\eqref{AppEq:dotSParm}) yields
\begin{equation}\label{AppEq:DS-Int-a}
\frac{\s_{,\z}}{\dot{\s}}-\mu_{,\z} \int^{\s}_0\frac{\dd \hs}{\hs (\dot{\hs})^3}=-t_{B,\z} \ .
\end{equation}
The integral above can be rewritten as
\begin{equation}\label{AppEq:Int-SdotSto3}
\int^{\s}_0\frac{\dd \hs}{\hs (\dot{\hs})^3}=\frac{1}{3 \mu} \left\{ k_0 \int^{\s}_0\frac{\dd \hs}{ (\dot{\hs})^3}+ \left(\frac{\dot{\s}}{\s}\right)^{-1}\right\} \ .
\end{equation}
Then, from~\eqref{AppEq:DS-Int-a} and~\eqref{AppEq:Int-SdotSto3}, we obtain:
\begin{equation}
\frac{\s_{,\z}}{\s}=
-t_{B,\z} \left(\frac{\dot{\s}}{\s}\right)+ 
k_0 \frac{\mu_{,\z}}{3 \mu} \left(\frac{\dot{\s}}{\s} \int^{\s}_0\frac{\dd \hs}{ (\dot{\hs})^3}\right)+ \frac{\mu_{,\z}}{3 \mu}  \ , 
\end{equation}
where we can identify the growing and decaying modes on the local background described by Eq.~\eqref{Eq:FriedmannLikeEqn}~\cite{PeacockBookCosPhys}:
\begin{equation}\label{EqApp:GrowDecS}
f_-=\frac{\dot{\s}}{\s} \ , \qquad f_+=\frac{\dot{\s}}{\s} \int^{\s}_0\frac{\dd \hs}{ (\dot{\hs})^3} \ .
\end{equation}
In addition to Eq.~\eqref{Eq:Ftt}, these functions satisfy
\begin{equation}\label{Eq:dfpmdt}
\s \,\dot{\s} \, \dot{f}_\pm -\left(k_0-\frac{3\mu}{\s}\right)\, f_\pm=\alpha_\pm \ ,
\end{equation}
\end{subequations}
with $\alpha_+=1$ and $\alpha_-=0$.

Finally, the metric coefficient $\GG$ can be cast into
\begin{equation}
\GG=f \frac{\s_{,\z}}{\s}+f \nu_{,\z}=\A-\left( \beta_+ f_+ +\beta_- f_-\right) \ ,
\end{equation}
where
\begin{subequations}
\begin{eqnarray}
\beta_+\equiv-f \frac{ k_0 \mu_{,\z}}{3 \mu} \ , \quad \beta_-\equiv f t_{B,\z} \ ; \label{Eq:betapmcI}
\\
 \A = f \nu_{,\z} -k_0 \beta_+= f_{,\z}-f\frac{\EE_{,\z}}{\EE}  -k_0 \beta_+\ .
\end{eqnarray}
\end{subequations}

Note that our definition of $\beta_-$ differs from the Goode and Wainwright choice by a multiplicative function constant in time, $6\mu$.

%%%%%%%%%%%%%%%%%%%%%%%%%%%%%%%%%%%%%%%%%%%%%%%%%%%%%%%%%%%%%%%
%%%%%%%%%%%%%%%%%%%%%%%%%%%%%%%%%%%%%%%%%%%%%%%%%%%%%%%%%%%%%%%
%%%%%%%%%%%%%%%%%%%%%%%%%%%%%%%%%%%%%%%%%%%%%%%%%%%%%%%%%%%%%%%

\section{GRZA model equations}
\label{SecApp:GRZAeqns}

Let us obtain the evolution equations for the deformation field. By direct substitution, we find that  
\begin{eqnarray}
&&\frac{1}{2 J} \epsilon_{abc} \epsilon^{ikl} \left( \dot{\eta}^a_{\ j} \eta^b_{\ k} \eta^c_{\ l} \right)^{\bdot}=
\frac{1}{\J} \left(\t{\ddot{\Pg}}{i}{j}+3 \hat{H} \t{\dot{\Pg}}{i}{j} \right)\qquad\qquad
\nonumber
\\
&&\qquad\qquad\qquad
+\left(\frac{\hat{H}\dot{\Pg}}{\J}+2 \hat{H}^2+\frac{\ddot A}{A} \right)\t{\delta}{i}{j} +\two\ ;
\\
&&\frac{1}{2J} \epsilon_{abc}\epsilon^{ik\ell} \ddot{\eta}^a_{\ i} \eta^b_{\ k} \eta^c_{\ \ell}   =
\frac{1}{\J}\left(\ddot \Pg+ 2\hat{H} \dot \Pg\right)+3\frac{\ddot A}{A} +\two 
 \ .
 \nonumber
 \\
 \label{Eq:ddn-n-nScal}
\end{eqnarray}
Using~\eqref{ricci} and~\eqref{raych} and keeping only the linear terms we obtain:
\begin{eqnarray}
&&-\CR^i_{\ j} =
\frac{1}{\J} \left(\t{\ddot{\Pg}}{i}{j}+3 \hat{H} \t{\dot{\Pg}}{i}{j} \right)
\nonumber
\\
&&\quad+\left[\frac{1}{\J}\left(\ddot \Pg+ 3\hat{H} \dot \Pg\right)
+4\frac{\ddot A}{A} + 2\left(\frac{\dot A}{A}\right)^2-2\Lambda\right]\t{\delta}{i}{j} 
 \nonumber
\\
&&\quad=\frac{1}{\J} \left(\t{\ddot{\Pg}}{i}{j}+3 \hat{H} \t{\dot{\Pg}}{i}{j} \right)+\left[\frac{1}{\J}\left(\ddot \Pg+ 3\hat{H} \dot \Pg\right)
-2\frac{\hat{k}}{A^2}\right]\t{\delta}{i}{j} \ .
\nonumber
\\
\label{AppEq:Rij-1}
\end{eqnarray}
Taking the trace leads to the evolution equation for $\Pg$:
\begin{equation}\label{EqApp:EvolTraceP}
\ddot{\Pg}+3 \hat{H} \dot{\Pg} = -\frac{\J}{4}\left(\CR-6\frac{\hat{k}}{\As^2}\right) \simeq -\frac{1}{4}\left(\CR-6\frac{\hat{k}}{\As^2}\right) \ .
\end{equation}
Note that $\CR-6\frac{\hat{k}}{\As^2}$ is of the order of $\Pg$. From now on (as in the main text), we will also assume that the Ricci tensor and $\hat{k}/\As^2$ are of the order of the deformation field as well. Substituting the above relation into Eq.~\eqref{AppEq:Rij-1} yields \eqref{SubEq:ddotPij}.
Multiplying this equation by the metric tensor $\mathbf{g}^{\One}$, contracting indices and neglecting the nonlinear terms, we get
\begin{equation}
\left(\ddot{\Pg}_{i j}+3 \hat{H} \dot{\Pg}_{i j} \right)=
-   \left[\frac{\CR_{i j}}{\As^2}-\frac{1}{4}\left(\CR+2\frac{\hat{k}}{\As^2}\right)G_{i j}\right] \ ,
\end{equation}
Next, we decompose the deformation field into the trace, tracefree symmetric, and antisymmetric parts, Eq.~\eqref{EqApp:DecompPij}.
Since $ \dot{\FP}_{i j}=0$, 
$\dot{\Pg}_{i j}=\frac{1}{3} \dot{\Pg} G_{ij} +\dot{\hat{\Pi}}_{i j}$ and 
$\ddot{\Pg}_{i j}=\frac{1}{3} \ddot{\Pg} G_{ij} +\ddot{\hat{\Pi}}_{i j}$, the evolution for the tracefree part simplifies to~\eqref{sympart_order1}.
For Szekeres models, this equation is equivalent to \eqref{EqApp:EvolTraceP}.

\subsection{Hamiltonian constraint}

Let us turn now to the Hamiltonian constraint; the left-hand term of Eq.~\eqref{form_hamiltoncoeff} can be expanded to give
\begin{equation}
\frac{1}{2J}\epsilon_{abc} \epsilon^{mjk} \dot{\eta}^a_{\ m} \dot{\eta}^b_{\ j} \eta^c_{\ k}=
3 \left(\frac{\dot A}{A}\right)^2 + 2 \hat{H} \frac{\dot \Pg}{\J} + \two \ . \qquad
\end{equation}
The density can be rewritten as
\begin{equation}
\varrho
=\frac{\varrho_{\ini}}{J}
=\frac{\hat{\varrho}_{b}}{A^3 \J}+\frac{\hat{\varrho}_{b}\hat{\delta}}{A^3 \J}=\frac{\hat{\varrho}_{b}}{A^3 \J}-\frac{\hat{W}}{4\pi A^3 \J}
\ ,
\end{equation}
with $ \hat{W}=-4 \pi \hat{\varrho}_{b} \, \hat{\delta}$. 
Finally, Eq.~\eqref{SubEq:dotP} results from substituting the previous expressions into~\eqref{form_hamiltoncoeff} and using~\eqref{Eq:FriedLikeA-1}.

\subsection{Connection coefficients}
\label{subsec:connection}

Since in this paper we only use the zero-order spatial Christoffel symbols,  we will not compute the higher order quantities. Injecting the zero-order spatial metric ($g^{\sm{(0)}}_{ij} = \As^2 G_{ij}$) and its inverse ($g^{\sm{(0)}ij} =  \As^{-2} G^{ij}$), 
into the definition of the Christoffel symbols,we obtain:
\begin{eqnarray}
&&{}^{\Zero}\Gamma^{i}_{\ kl} =
\ToZero\overline{\Gamma}^{i}_{\ kl}  + \ToZero\widetilde{\Gamma}^{i}_{\ kl}  ,
\qquad \hbox{with}
\\
 && 
{}^{\Zero} \overline{\Gamma}^{i}_{\ kl}=  \frac{A_{| l}}{A} \t{\delta}{i}{k} + \frac{A_{| k}}{A} \t{\delta}{i}{l} -   \frac{A_{| m}}{A} G^{i m} G_{ k l}  \ ;
\\
&&
{}^{\Zero} \widetilde{\Gamma}^{i}_{\ kl}=\frac{G^{ im}}{2}  \left(G_{ m k |l} + G_{ l m | k} - G_{ k l | m} \right)  \ .
\label{EqApp:ChristGij}
\end{eqnarray}
Note that  ${}^{\Zero} \widetilde{\Gamma}^{i}_{\ kl}$ are the Christoffel
symbols for $G_{ij}$: 
\begin{equation}
{}^{\Zero} \widetilde{\Gamma}^{i}_{\ kl}=\t{\Gamma}{i}{j k}(\mathbf{G}) \ .
\end{equation}

In the context of large-scale structure formation, it is convenient to assume that the initial metric is close to the flat FLRW line-element in Cartesian coordinates:
\begin{equation}\label{Eq:SplittingFLRW}
G_{i j}\simeq \delta_{i j} + \delta G_{i j} \ ,
\end{equation}
where the quantities $\delta G_{i j}$ are small and of the order of magnitude of the deformation field. Note that, in general, the Szekeres metric does not admit a splitting of the form~\eqref{Eq:SplittingFLRW}. This is due to the fact that, in Szekeres models, 
FLRW emerges in an unfamiliar coordinate system. 
In this work, we have adopted a general Gram's matrix to represent the initial metric,
which allows for recovering the complete Szekeres family under a proper selection of the initial data.

%%%%%%%%%%%%%%%%%%%%%%%%%%%%%%%%%%%%%%%%%%%%%%%%%%%%%%%%%%%%%%%%%%%%%%%%%%
%%%%%%%%%%%%%%%%%%%%%%%%%%%%%%%%%%%%%%%%%%%%%%%%%%%%%%%%%%%%%%%%%%%%%%%%%%
%%%%%%%%%%%%%%%%%%%%%%%%%%%%%%%%%%%%%%%%%%%%%%%%%%%%%%%%%%%%%%%%%%%%%%%%%%
\subsection{Momentum constraints}

Let us consider the momentum constraints, Eq.~\eqref{form_momcoeff}, and note that
\begin{eqnarray}
\t{\lambda}{i}{j}&\equiv&\frac{1}{J}\epsilon_{abc} \epsilon^{ikl} \dot{\eta}^a_{\ j} \eta^b_{\ k} \eta^c_{\ l} = 2 \left(\hat{H}\t{\delta}{i}{j}+\t{\dot{\Pg}}{i}{j}/\J\right) + \two \ ;
\nonumber
\\
\\
\hat{\lambda}&\equiv&\frac{1}{J}\epsilon_{abc} \epsilon^{ikl} \dot{\eta}^a_{\ i} \eta^b_{\ k} \eta^c_{\ l} = 6 \hat{H} +2 \dot{\Pg}/\J + \two \ .
\end{eqnarray}
Then, let us compute each term of the covariant derivative 
($\t{\lambda}{i}{j||i}=\t{\lambda}{i}{j|i} + \t{\Gamma}{i}{i l} \t{\lambda}{l}{j} -  \t{\Gamma}{l}{i j} \t{\lambda}{i}{l}$) on the left-hand side of Eqs.~\eqref{form_momcoeff},
\begin{eqnarray}
\t{\lambda}{i}{j|i} &=&2\frac{\dot{\As}_{|j}}{\As}-2 \hat{H} \frac{\As_{|j}}{\As}+ 2 \frac{\t{\dot{\Pg}}{i}{j|i}}{\J} +\two \ ; 
\label{AppEq:dlamjDisc}
\\
\t{\Gamma}{i}{i l} \t{\lambda}{l}{j} &-&  \t{\Gamma}{l}{i j} \t{\lambda}{i}{l} \nonumber
\\
&=& 2 \hat{H} \left(\t{\Gamma}{i}{i l}\t{\delta}{l}{j}-\t{\Gamma}{l}{i j} \t{\delta}{i}{l}\right) 
+\frac{2}{\J}\left(\t{\Gamma}{i}{i l}\t{\dot{\Pg}}{l}{j}-\t{\Gamma}{l}{i j} \t{\dot{\Pg}}{i}{l}\right) 
\nonumber
\\
&=& 
 \frac{2}{\J} \left(\t{\ToZero\Gamma}{i}{i l}\t{\dot{\Pg}}{l}{j}-\t{\ToZero\Gamma}{l}{i j} \t{\dot{\Pg}}{i}{l}\right) + \two\ . \quad
\end{eqnarray}
Putting all these results together, we obtain:
\begin{eqnarray}
&&\left(\tfrac{1}{J}\epsilon_{abc} \epsilon^{ikl} \dot{\eta}^a_{\ j} \eta^b_{\ k} \eta^c_{\ l} \right)_{||i}
=2\frac{\dot{\As}_{|j}}{\As}-2 \hat{H} \frac{\As_{|j}}{\As}
\nonumber
\\
&&
\qquad\qquad\qquad\qquad
+ 2 \frac{\t{\dot{\Pg}}{i}{j|i}}{\J}- 4\zeta_j +\two \ ;
\\
&&\left(\tfrac{1}{J}\epsilon_{abc} \epsilon^{ikl} \dot{\eta}^a_{\ i} \eta^b_{\ k} \eta^c_{\ l} \right)_{|j} =
6\frac{\dot{\As}_{|j}}{\As}-6 \hat{H} \frac{\As_{|j}}{\As} 
\nonumber
\\
&&\qquad\qquad\qquad\qquad\qquad+ 2\frac{\t{\dot{\Pg}}{i}{i|j}}{\J} +\two \ ,
\end{eqnarray}
where $ \zeta_j\equiv-\frac{1}{2\J}\left(\t{\ToZero\Gamma}{i}{i l}\t{\dot{\Pg}}{l}{j}-\t{\ToZero\Gamma}{l}{i j} \t{\dot{\Pg}}{i}{l}\right)$; then, we obtain \eqref{Eq:momConstSys1}, 
where $\zeta_j$ can be written out to give Eq.~\eqref{Eq:zeta_j}.

Using~\eqref{EqApp:DecompPij}, the term $\t{\dot{\Pg}}{i}{[i|j]}$ can be rewritten as follows:
\begin{eqnarray}
\t{\dot{\Pg}}{i}{[i|j]}&&=\frac{1}{2} \left(\t{\dot{\Pg}}{i}{i|j}-\t{\dot{\Pg}}{i}{j|i}\right)
=\frac{1}{2} \left(\dot{\Pg}_{|j}-\t{\dot{\Pg}}{i}{j|i}\right)
\\
&&=\frac{1}{2} \left[\dot{\Pg}_{|j}-\left(\frac{1}{3} \dot{\Pg} \t{\delta}{i}{j}+\t{\dot{\hat{\Pi}}}{i}{j}\right)_{|i}\right]
=\frac{1}{3} \dot{\Pg}_{|j}-\frac{1}{2} \t{\dot{\hat{\Pi}}}{i}{j|i} \ .
\nonumber
\\
\end{eqnarray}
Finally, we obtain the linear expression for the momentum constraints, Eq.~\eqref{momentum_order1}.
Under the assumption~\eqref{Eq:SplittingFLRW}, $\zeta_j$ simplifies to
\begin{equation}
     \zeta_j=\frac{1}{2}\left(
\frac{A_{| j}}{A} \dot{\Pg}
-3 \frac{A_{| l}}{A}\t{\dot{\Pg}}{l}{j} \right) \; .
\end{equation}
%

%%%%%%%%%%%%%%%%%%%%%%%%%%%%%%%%%%%%%%%%%%%%%%%%%%%%%%%%%%%%%%%
%%%%%%%%%%%%%%%%%%%%%%%%%%%%%%%%%%%%%%%%%%%%%%%%%%%%%%%%%%%%%%%
%%%%%%%%%%%%%%%%%%%%%%%%%%%%%%%%%%%%%%%%%%%%%%%%%%%%%%%%%%%%%%%

\section{Backreaction and emergence of spatial curvature}
\label{SecApp:EmergingCurvature}

Consider a compact domain ${\cD}$, the volume scale factor of which being defined as follows:
\begin{equation}
a_{\cD}(t)=\left(V_{\cD}/V_{\cD_i}\right)^{\frac{1}{3}} \ ; \quad V_{\cD_i}=V_{\cD}(t_i) \ .
\end{equation}
Then, $H_\cD=\dot{a}_\cD/a_\cD$ represents the volume Hubble expansion rate.
 
In a general spacetime, the volume expansion and volume acceleration generalize the kinematic laws of a homogeneous and isotropic FLRW spacetime. In this framework, emerging curvature is a result of its coupling to the kinematical backreaction term (involving  the first two principal scalar invariants of the expansion tensor) via Eq. \eqref{AppEq:Integrability} below \cite{Buchert2000GERG,BuchertFocus}:  
\begin{subequations}\label{Eqs:BackreactionSys}
\begin{eqnarray}
\CQ_{\cD}&\equiv& 2 \average{\inII}-\frac{2}{3} \average{\inI}^2  \label{Eq:QdInv-a}
\\
&=& \frac{2}{3} \left( \average{\Theta^2} - \average{\Theta}^2\right)-2 \average{\sigma^2} \ . \label{Eq:QdScal-b}
\end{eqnarray}
\end{subequations}
In these equations, $\sigma^2 \equiv \frac{1}{2}\t{\sigma}{a}{b}\t{\sigma}{b}{a}$ while $\inI$ and $\inII$ (and $\inIII$) represent the principal scalars invariant of the expansion tensor, $\inI=\t{\Theta}{i}{i}=\Theta$, $\inII=\frac{1}{2}\left(\Theta^2-\Theta^i_{\ j}\Theta^j_{\ i}\right)=\frac{1}{3} \Theta^2-\sigma^2$, which are defined as follows for an arbitrary matrix $\t{B}{i}{j}$
\begin{subequations}\label{SubEqs:DefInvBij}
\begin{eqnarray}
\inI(\t{B}{i}{j})&=&\frac{1}{2} \t{\epsilon}{-}{abc} \t{\epsilon}{ikl}{} \t{B}{a}{i} \t{\delta}{b}{j} \t{\delta}{c}{k} \ ;
\\
\inII(\t{B}{i}{j})&=& \frac{1}{2} \t{\epsilon}{-}{abc} \t{\epsilon}{ikl}{} \t{B}{a}{i} \t{B}{b}{j} \t{\delta}{c}{k} \ ;
\\
\inIII(\t{B}{i}{j})&=& \frac{1}{6} \t{\epsilon}{-}{abc} \t{\epsilon}{ikl}{} \t{B}{a}{i} \t{B}{b}{j} \t{B}{c}{k} \ .
\end{eqnarray}
\end{subequations}
Introducing the deviation from a (scale-dependent) constant curvature model as a new backreaction variable,
$\CW_\cD : = \average{\CR} - 6 k_{\cD_{\rm i}} / a_\cD^2$, 
the volume-averaged set of equations reads \cite{Buchert2000GERG}: 
\bse\label{AppEq:SetAvedEqsQd}
\begin{eqnarray}
\left(\frac{\dot{a}_\cD}{a_\cD}\right)^2 -\frac{\Lambda}{3} - \frac{8 \pi}{3} \average{\varrho} + \frac{k_{\cD_{\rm i}}}{a_\cD^2} = 
- \frac{1}{6} ( \CQ_\cD +  \CW_\cD )  \, ; \qquad
\label{AppEq:BackR-da}
\\
\frac{\ddot{a}_\cD}{a_\cD}- \frac{\Lambda}{3} + \frac{4 \pi}{3} \average{\varrho} = \frac{1}{3}\CQ_{\cD} \, ;  \qquad\label{AppEq:BackR-dda}
\\
\dotaverage{\varrho} + 3 \frac{\dot{a}_\cD}{a_\cD} \average{\varrho} = 0\, ; \qquad\label{AppEq:BackR-rho}
\\
\frac{1}{a_\cD^6} \left(\CQ_{\cD} a_\cD^6\right)^{\bdot} + \frac{1}{a_\cD^2} \left (\CW_\cD a_\cD^2 \right)^{\bdot} =0 ,\qquad \label{AppEq:Integrability}
\end{eqnarray}
\ese
where~\eqref{AppEq:BackR-dda}, \eqref{AppEq:BackR-rho} and~\eqref{AppEq:BackR-rho} correspond to the  
averaged energy constraint, Raychaudhuri's and rest mass conservation equations, respectively. Equation~\eqref{AppEq:Integrability} is not independent; it emerges as a condition of integrability in order to obtain~\eqref{AppEq:BackR-dda} through the time-derivative of~\eqref{AppEq:BackR-da}.

%%%%%%%%%%%%%%%%%%%%%%%%%%%%%%%%%%%%%%%%%%%%%%%%%%%%%%%%%%%%%%%
%%%%%%%%%%%%%%%%%%%%%%%%%%%%%%%%%%%%%%%%%%%%%%%%%%%%%%%%%%%%%%%
%%%%%%%%%%%%%%%%%%%%%%%%%%%%%%%%%%%%%%%%%%%%%%%%%%%%%%%%%%%%%%%   

\section{Numerical example in the language of exact solutions}
\label{App:NumExSol}

The initial data of the family of models examined in Section \ref{Sec:ExactbodyNum} is set up in the Hellaby's parametrization, where the physical interpretation of the variables is clearer than in the GW's. The initial areal distance is taken equal to the `radial' comoving coordinate 
\begin{equation}
\Phi(t_{\ini},\z)= \z \ . 
\end{equation}
The solution of interest is contained in the domain $\cD: 0\leq \z \leq \z_1, -\infty < x,y < +\infty$. Here, $\z_1$ sets the length scale of the initial configuration. 

The free function $M(\z)$ is chosen as
\begin{equation}
M(\z)=\frac{4\pi}{3}\varrho_{bi} \z^3 \ ,
\end{equation}
where $\varrho_{bi}$ is the density of the homogeneous universe at the initial time (i.e., the last scattering time) and $\hat{k}(\z)$ is specified as in Eq.~\eqref{Eq:hatk}.

Then, we introduce the dimensionless variables \eqref{Eq:NumExNoDimPar} using the Hubble parameter at the last scattering time as the timescale and the characteristic length  $\ell=2\times 10^{-2}$. (This characteristic length expands to supercluster scales at present time.) This solution is rewritten in the GW language using the results of Appendix \ref{SubSec:RelHella-GW}, and then rescaled as sketched in Section~\ref{SubSec:NormlChi}. In terms of the Szekeres functions and the perturbation, the line-element reads: 
\begin{eqnarray}
\dd s^2=-\dd t^2 + \As^2  \bigg[ \left(\A-\hat{\F}\right)^2 \left(\frac{\W}{\chi}\right)^2 \dd \zd^{2} 
\nonumber
\\
+ \zd^{2} \E^{-2}  \left( \dd \x^{2} + \dd \y^{2}\right) \bigg] \ ,
\label{AppEq:MetricNumEx}
\end{eqnarray}
where we have used that $\E$ was defined in \eqref{Eq:eToNuGW}.

Inserting Eq.~\eqref{AppEq:MetricNumEx} into \eqref{Eq:HamEst}, taking back the rescaling of the solution and using the Szekeres exact relations, we obtain the following expression for the violation of the Hamiltonian constraint: 
\begin{eqnarray}
\mathbb{H}(\rv,t):=
\frac{2 }{\As^2 \zd^2 \W^2}
\bigg[
-\frac{1}{2}\left(1+\frac{\GG}{\hat{\GG}}\right)
-\frac{1}{\hat{\GG}} f \nu_{,\zd}
\nonumber
\\
+\epsilon f_{,\zd}
+\frac{f \GG_{,\zd}}{ \hat{\GG}^2}
\bigg]\left(1-\frac{\GG}{\hat{\GG}}\right)
\nonumber
\\
-\frac{\E^2}{\As^2 \zd^2 \hat{\GG}}
\bigg[\left(k_0+f_+\right)\left(\delta\beta_{+,\x\x}+\delta\beta_{+,\y\y}\right)
\nonumber
\\
+f_-\left(\delta\beta_{-,\x\x}+\delta\beta_{-,\y\y}\right)\bigg]
\nonumber
\\
+\frac{2 f}{ \As^2 \zd^2 \W^2 \hat{\GG}^2} 
\left(\frac{\GG}{\hat{\GG}}\right)\bigg[\delta \beta_+(k_0+f_+)+ \delta \beta_- f_- \bigg]_{,\zd}.
\quad
\label{Eq:HamError}
\end{eqnarray}
\end{appendix}

%%%%%%%%%%%%%%%%%%%%%%%%%%%%%%%%%%%%%%%%%%%%%%%%%%%%%%%%%%%%%%%
%%%%%%%%%%%%%%%%%%%%%%%%%%%%%%%%%%%%%%%%%%%%%%%%%%%%%%%%%%%%%%%
%%%%%%%%%%%%%%%%%%%%%%%%%%%%%%%%%%%%%%%%%%%%%%%%%%%%%%%%%%%%%%%   
\newcommand\eprintarXiv[1]{\href{http://arXiv.org/abs/#1}{arXiv:#1}}
%\section*{References}                
%\bibliographystyle{plain}
%\bibliography{SDcites}

%
\end{document}